%% file: ecrc-template.tex
\documentclass[preprint,3p]{elsarticle}

\usepackage[T1]{fontenc}

\usepackage{graphicx}
\usepackage{subcaption}

\usepackage{makecell}

\usepackage{longtable}

\usepackage{multirow}

\usepackage{xcolor}
\usepackage{listings}

\usepackage[hidelinks]{hyperref}

\usepackage{amssymb}

\usepackage{color}
\usepackage{listings}
\usepackage[figuresright]{rotating}

\journal{Computer Networks}

\begin{document}

\begin{frontmatter}


\author{Alexandre Gustavo Wermann\corref{cor1}\fnref{label1}}
\ead{agwermann@inf.ufrgs.br}
\author{Juliano Araujo Wickboldt\fnref{label1}}
\ead{jwickboldt@inf.ufrgs.br}
\ead[url]{http://www.inf.ufrgs.br/~jwickboldt}
\cortext[cor1]{Corresponding author}
\address[label1]{Federal University of Rio Grande do Sul (UFRGS) -- Institute of Informatics\\Av. Bento Gonçalvez, 9500, Porto Alegre, RS, 91.509-900, Brazil}

 
\title{KTWIN: A Serverless Kubernetes-based Digital Twin Platform}




\include{0-abstract}


\end{frontmatter}



\input{1-introduction}
\input{2-literature-review}
\input{3-proposal}
\input{4-implementation}
\input{5-evaluation}
\input{6-conclusion}





\bibliographystyle{elsarticle-num}
\bibliography{biblio}







\end{document}

%% file: 0-abstract.tex
\begin{abstract}
Digital Twins (\textit{DTs}) systems are virtual representations of physical assets allowing organizations to gain insights and improve existing processes. In practice, \textit{DTs} require proper modeling, coherent development and seamless deployment along cloud and edge landscapes relying on established patterns to reduce operational costs. In this work, we propose \textit{KTWIN} a Kubernetes-based Serverless Platform for Digital Twins. \textit{KTWIN} was developed using the state-of-the-art open-source Cloud Native tools, allowing \textit{DT} operators to easily define models through open standards and configure details of the underlying services and infrastructure. The experiments carried out with the developed prototype show that \textit{KTWIN} can provide a higher level of abstraction to model and deploy a Digital Twin use case without compromising the solution scalability. The tests performed also show cost savings ranging between 60\% and 80\% compared to overprovisioned scenarios.
\end{abstract}

%% file: 1-introduction.tex
\section{Introduction}
\label{section:introduction}

Digital Twins (\textit{DTs}) are dynamic virtual model representations of real-world objects, processes, or services. These systems are designed to collect highly distributed data from different devices, sensors, and other sources to create a reliable virtual model of a physical asset or process \cite{dt-model-based-systems-engineering}. The data collected enable organizations to monitor performance, predict potential issues, and implement responsive measures using advanced analytics and Artificial Intelligence-based models. The adoption of Digital Twins spans various industries such as manufacturing, health care, smart cities, and the energy sector.

The growing number of connected devices and the increasing volume of data generated make it challenging to move and process all the data in the Cloud, especially for time-sensitive application requirements. The advances of Mobile Edge Computing (\textit{MEC}) technologies drove the deployment of micro data centers closer to end users to fit time requirements and, at the same time, overcome computing resource constraints and energy consumption restrictions of the Internet of Things \textit{IoT} edge devices \cite{mec-offloading-edge-devices}. In addition, Mobile Cloud Computing (\textit{MCC}) architecture emerges as a solution for offloading data processing from edge devices to centralized cloud data centers, thus enhancing the user experience and providing greater scalability with access to extensive cloud resources \cite{offloading-mobile-edge-cloud-computing}. In such a complex and diverse landscape, building, deploying, and operating Digital Twin components becomes difficult; hence, it is important to establish operations patterns and rely on automation and abstractions to reduce operational burden and costs.

Building a \textit{DT} system requires a solid understanding of its specific domain requirements to design the diverse number of entities and their relationships. There are several ways to define Digital Twin entities using Ontology-based models, such as \textit{OntoUML} and \textit{DEXPI}. More recently, the open-source community joint effort with industry has designed the Digital Twin Definition Language (\textit{DTDL}) \cite{dtdl-spec} which allows developers and domain experts to describe the twin graph in self-defined terms of domain models. However, since these languages focus on \textit{DT} model definitions, there is a lack of ways to describe technical application aspects that compose the deployment of a Digital Twin, such as the number of allocated CPU and memory, the software implementation required to respond to specific events and the application auto-scaling settings.

Moreover, existing platforms are mostly vendor-specific implementations in which end-users cannot easily move their solution to another vendor stack without a considerable amount of rework, causing the vendor lock-in problem. Additionally, because the existing offerings operate mostly in the cloud, there is not enough flexibility to move services to edge locations, closer to the end user. An automated vendor-agnostic and unified edge-to-cloud platform solution could reduce the operational burden while providing a higher level of abstraction to keep \textit{DT} services up and running.

Hence, in this research, we propose \textit{KTWIN}, a Serverless Kubernetes-based platform, that allows \textit{DT} operators to easily define models through open standards and configure details of the underlying service and infrastructure, such as resource allocation, container-based services deployments, and auto-scaling policies. {\color{black}\textit{KTWIN} utilizes Serverless principles to abstract infrastructure management, enabling on-demand scaling, efficient resource utilization, and reduced operational overhead while ensuring high availability and responsiveness for Digital Twin applications.} The platform was designed using the state-of-the-art Cloud Native open-source technologies, and it can be deployed in any Kubernetes cluster, in the cloud, or at edge locations. \textit{KTWIN} leverages the rich environment of Kubernetes and allows for further enhancements and open collaboration within the \textit{DT} industry and academy. Additionally, \textit{KTWIN} maps and orchestrates domain entity specification to containers with implementations that can be defined by the \textit{DT} owners. The benefits of using \textit{KTWIN} include: (1) reduction of operational costs by automating manual operations, (2) more control and flexibility over the underlying container-based services that compose the \textit{DT} landscape, (3) can be deployed at any Kubernetes cluster (on-premise, in the edge, in the cloud) reducing the chance of vendor lock-in.

The main contributions of this research are (1) the definition of an extended \textit{DTDL} language to provide a higher level of abstraction in which \textit{DT} owners can define domain entities and at the same time have control over the underlying services and operational layer, (2) the design and implementation of a Kubernetes Operator to orchestrate the \textit{DT} components as containers as well as all the underlying services, messaging routing rules and data storage, (3) and the deployment of a New York City-based Smart City Digital Twin use-case to validate the usage of the platform. The verified prototype results show that it is possible to provide a higher level of abstraction and model design flexibility while still providing scalability and good system performance. The experiments also compare \textit{KTWIN} resource usage and response time against scenarios of under-provisioned and over-provisioned resources. \textit{KTWIN} achieves resource usage efficiency closer to the under-provisioned scenario, while still providing performance near the over-provisioned scenario.

The remainder of this article is organized as follows. Section \ref{section:related-work} presents previous work in the \textit{DT} field including real-world use cases, aspects regarding modeling and designing \textit{DTs} and the existing cloud offerings. Section \ref{section:ktwin-architecture-requirements-design} presents the \textit{KTWIN} requirements and its architecture design including their main components, resource definition, message routing, and event storage. Section \ref{section:ktwin-implementation} explains the implementation details of the proof-of-concept built to evaluate the proposed design. In Section \ref{section:ktwin-evaluation-results}, we present the evaluation methodology and discuss the results of the proposed experiments. Finally, Section \ref{section:conclusion} concludes the research and present future works.

%% file: 2-literature-review.tex
\section{Background}
\label{section:background}

In recent years, the Digital Twin (\textit{DT}) research field has seen significant advancements and contributions from academia leveraging the benefits of \textit{DT} systems to several industries. Researchers say that the Digital Twin market size can be worth US\$ 110.1 billion by 2028 \cite{dt-markets}. This section reviews the different use cases that Digital Twins can be applied, as well as the existing platforms and frameworks offerings and their limitations.

\subsection{Digital Twin Use Cases}

A Digital Twin is a virtual representation of a physical system or process, designed to facilitate bidirectional communication between the real-world system and its digital counterpart \cite{Lehner-2022}. \textit{DTs} enable the development of value-added services, such as real-time monitoring, predictive maintenance, optimization, and simulation, that can be adopted by a variety of industries such as Manufacturing, Smart Cities, Energy and Healthcare.

In the context of the Manufacturing area, a \textit{DT} can be utilized to solve several field problems such as optimizing production processes, monitoring equipment health, applying optimization to the supply chain process, and performing predictive maintenance activities. Sierla \textit{et al.} \cite{automatic-assembly-SIERLA201834} build a Digital Twin to automate assembly planning and orchestrate the production resources in a manufacturing cell based on digital product descriptions. The concept introduced in the research is presented in general terms in \textit{UML} (Unified Modeling Language) and the implementation provides a 3D simulation environment using Automation Markup Language for digital product descriptions. Tao an Zhang \cite{shop-flow-tao-8049520} propose a Digital Twin Shop-floor (\textit{DTS}) providing an effective way to reach the physical-virtual convergence in the manufacturing industry and improve production efficiency, accuracy, and transparency. The system components are broken down into a Physical Shop-floor (\textit{PS}), Virtual Shop-floor (\textit{VS}), Shop-floor Service System (\textit{SSS}), and Shop-floor Digital Twin Data (\textit{SDTD}). The paper also presents the key technologies required to build the proposed system and the challenges during a \textit{DTS} implementation process for future studies. It lacks an actual implementation and evaluation of the proposed solution. 

The modernization of cities opens the adoption of Digital Twin technologies to improve problems of medium-sized cities around the globe. Digital Twins can be applied to improve public transportation systems and address mobility problems, enhance the decision-making process in urban planning, monitor and predict environmental and resource management such as carbon emission and waste reduction, or detect anomalies and optimize operations in power and water plants. Dembski \textit{et al.} \cite{dembski2020urban} present a novel Digital Twin case study applied in the town of Herrenberg, Germany. The research presents a prototype and the deployment of sensor devices in different city locations to collect data. The use case prototype includes a 3D city model, a mathematical street network model, urban mobility simulation, airflow simulation based on collected environment data, sensor network data analysis and people's movement routes. The prototype allows developers to easily extend the urban digital flow by adding new code and modules in a visual interface. Ford and Wolf \cite{ford2020smart} exploit Smart Cities with Digital Twins (\textit{SCDT}) for Disaster Management Digital Twin system for Smart Cities. The authors present and test a conceptual model of a \textit{SCDT} for disaster management and discuss the issues to be addressed in the development and deployment of \textit{SCDT} for disaster management.

Digital Twins can improve the energy industry by providing valuable insights into system performance and optimization, reducing power downtime, providing efficient energy planning to fit the supply and demand, as well as optimizing energy household consumption. In this context, Fathy and Jaber \cite{energy-consumption-planning-fathy2021digital} propose a data-driven multi-layer Digital Twin of the energy system that aims to mirror the actual energy consumption of households in the form of a Household Digital Twin (\textit{HDT}). The model intends to improve the efficiency of energy production, modeled as an Energy Production Digital Twin (\textit{EDT}), by flattening the daily energy demand levels. This is done by collaboratively reorganizing the energy consumption patterns of residential homes to avoid peak demands while accommodating the resident's needs and reducing their energy costs. In addition, Gao \textit{et al.} \cite{dt-energy-containers-gao2023digital} present a Digital Twin-based approach to optimize the operation of an automatic stacking crane handling containers regarding energy consumption. The article developed a virtual container yard that syncs real container yard information to improve automatic stacking crane scheduling. Then, a mathematical model followed by a Q-learning-based optimization step is applied to minimize the total energy consumption for completing all tasks. The output can be used by managers and operators to choose the appropriate strategy to accomplish energy-saving and efficiency goals. In the Healthcare ecosystem, Digital Twins can be applied for many use cases such as creating a model replica of a single patient for health monitoring and treatment planning, simulating and testing a medical device during the design phase or even modeling an entire healthcare system to monitor to analyze vaccination campaigns, for instance. Pilati \textit{et al.} \cite{pilati2021digital} developed a Digital Twin which integrates the physical and virtual health systems to create a sustainable and dynamic vaccination center focusing on using the minimum space and resources while guaranteeing a good service for patients. The research implements a discrete event simulation model to try to find the best configuration in terms of human resources and queues for a real vaccination clinic to maximize the number of patients vaccinated in the shortest period. 

In the context of health monitoring and treatment planning, Liu and Zhang \textit{et al.} \cite{liu-healthcare-cloud-dt} propose a novel cloud-based Digital Twin healthcare framework for elderly patients (CloudDTH). The solution plans, monitors, diagnoses and predicts the health of individuals by data acquired via wearable medical devices, toward the goal of personal health management. The proposed implementation can be divided into three phases: crisis early warning, real-time supervision, and scheduling and optimization. The data utilized to monitor and simulate individual health includes data acquired from physical medical wearable devices from elderly patients (e.g. heartbeat, body temperature, blood pressure), external environmental factors such as temperature change, and medical records.

\subsection{Digital Twins: Existing Services and Frameworks}

In the general context of Digital Twin platforms Eclipse Ditto \cite{eclipse-ditto}, Microsoft Azure Digital Twins \cite{azure-digital-twin} and AWS IoT TwinMaker \cite{aws-twin-maker} are the primary references. The Azure Digital Twin is a Platform as a Service (\textit{PaaS}) offered by the Microsoft Azure suite that enables organizations to create digital replicas of physical environments, systems, and processes. Microsoft's Digital Twin platform leverages the power of the Azure cloud and integrates with a wide range of other Microsoft services and tools, such as IoT Hub, Event Hubs, Event Grids, and Service Bus. Azure Digital Twins service allows to model entities and their relationship in a JSON-like language called Digital Twin Definition Language (\textit{DTDL}) forming a conceptual graph that can be visualized and queried in the Digital Twin Explorer.

Amazon Web Services offers the AWS \textit{IoT} TwinMaker service to build operational digital twins of physical and digital systems. AWS \textit{IoT} TwinMaker allows organizations to create digital visualizations using measurements and analysis from a variety of real-world sensors, cameras, and enterprise applications to help you keep track of your physical factory, building, or industrial plant. The platform also provides tools to model your system by using an entity-component-based knowledge graph composed of entities, components, and relationships using a proprietary JSON-like language. In addition, the TwinMaker service allows configuring and loading data from time-series data sources, and it builds 3D visualizations on top of it.

Eclipse Ditto is an open-source framework for Digital Twins maintained by Eclipse Foundation. Eclipse Ditto focuses on connecting physical devices to their digital counterparts, enabling interaction and synchronization between both. It allows device virtual representation encompassing their state, properties and behaviors. Ditto supports multiple protocols for interacting with digital twins, allowing users to query, modify, and subscribe to changes in the twin's state.

The existing PaaS solutions offer a wide variety of features, but they do not provide enough flexibility to deploy Digital Twin solutions to different landscapes, such as edge locations, and limit organizations to work with hybrid or multi-cloud environments. Moreover, they mostly are vendor-specific implementations, reducing interoperability and reusability \cite{Lehner-2022}, and limiting organizations to move their solutions to another vendor stack without a considerable amount of rework. Lastly, all the services mentioned above offer less control over operations and infrastructure, preventing users from personalizing settings, and monitoring application health and events propagation. In this context, an open-source Digital Twin platform built with Cloud Native solutions can offer flexibility and independence to organizations to build microservice-based distributed systems to leverage more Digital Twin use cases.

\section{Previous Work in Digital Twins}
\label{section:related-work}

{\color{black}

This section delves into the state of the art in \textit{DT} research, highlighting recent advancements, frameworks, and architectures, in the field. Grieves \textit{et al.}\cite{Grieves2017} formalized that a \textit{DT} system is a composition of three primary elements: a real-world space composed of a set of objects or systems, a virtual space representation of the real-world space, and a link in which the data flow between both spaces. Tao \textit{et al.} \cite{Tao2018} enhanced the definition and established the Five-Dimensional Digital Twin model composed by: a physical entity model, virtual entity model, services model, data model and connection model.

}


Several prior works have proposed frameworks for implementing \textit{DTs)}. Talasila \textit{et al.} \cite{talasila-digital-twin-dtaas} propose a Digital Twin framework to manage Digital Twin assets, making them available as a service to other users. The proposed framework automates the management of reusable assets, storage, provision of computing infrastructure, communication and monitoring tasks. Users operate at the level of Digital Twins and delegate the rest of the work to the Digital Twin as a service framework. The paper does not present any real implementation or evaluation scenarios, relying on high-level design aspects of what a Digital Twin platform should contain.

Picone \textit{et al.} article \cite{picone-edge-digital-twin} presents the Edge Digital Twins (\textit{EDT}) architectural model and its implementation. \textit{EDT} enables lightweight replication of physical devices and provides a digital abstraction layer to facilitate the interaction with \textit{IoT} devices and services, enhancing \textit{IoT} digitalization and interoperability. The EDT architectural blackprint targets to be complementary rather than in competition with existing edge modules and DT solutions (also in the cloud). The proposal provides the following advantages: remove the responsibility from physical assets to handle the integration with cloud \textit{DTs}, reduce vendor lock-in with cloud providers by introducing \textit{EDTs} as middleware layers between devices and cloud \textit{DT} services, and allow applications to interact with \textit{EDT} locally with MQTT protocol improving performance and reliability.

Furthermore, Wang \textit{et al.} \cite{wang-mobility-digital-twin} developed a Mobility Digital Twin (\textit{MDT}) framework defined as an artificial intelligence-based data-driven cloud–edge–device framework for mobility services. The \textit{MDT} consists of three static building blocks in the physical space (Human, Vehicle, and Traffic) and their associated Digital Twins in the digital space. The paper presents an example of cloud–edge architecture built with Amazon Web Services (AWS) to accommodate the proposed MDT framework and to fulfill its digital functionalities of storage, modeling, learning, simulation, and prediction.

{\color{black}

Context-aware and adaptive Digital Twin models have been studied by researches. Hribernik \textit{et al.} \cite{Hribernik-2021} discusses the potential of context-aware, autonomous, and adaptive \textit{DTs} being the building blocks for tomorrow's Digital Factories. Bellavista \textit{et al.} \cite{Bellavista-2023} propose a vision of \textit{DT} design and implementation to contain adaptive, autonomous, and context-aware functionalities. The paper also propose and implements a set of reusable micro-services design patterns allowing engineers to meet these new demanding requirements while keeping complexity and management costs under control.

Minerva \textit{et al.} \cite{Minerva-2020}  provide a comprehensive survey of technologies, scenarios, application cases, and architectural models relevant to the implementation of the \textit{DT} concept. The study consolidates key definitions, specifications, and implementations of \textit{DTs} across diverse technological domains, offering insight into how the concept has evolved and been adopted in various industries. The article explores the applicability of \textit{DTs} in key \textit{IoT} application scenarios and proposes a framework for evaluating its implementation with modern software architectures.

Bicocchi \textit{et al.} \cite{bicocchi-dt-requirements} explored the concept of end-to-end trustworthiness in the context of \textit{DTs}. The research identifies the six key characteristics that enable trustworthiness, i.e., entanglement awareness, variable load resilience, edge-to-cloud continuum mobility, declarative representation, observability, and security. In addition, The authors designed a blackprint architecture that natively supports end-to-end trustworthiness and discussed a proof-of-concept implementation.

}


In the context of serverless computing, Wang \textit{et al.} \cite{supporting-iot-applications-serverless} discuss the usage of a Serverless architecture for \textit{IoT} applications and build a prototype with Cloud-Native solutions to be deployed at the edge. The authors state that according to their experiments, the serverless approach uses fewer resources than the traditional ``serverfull'' computing approach. In contrast, serverless workloads presented a higher response time due to cold-start start-up time. The prototype and its evaluation were not implemented considering the Digital Twin context, focusing on a more general serverless \textit{IoT} use case.

Bellavista \textit{et al.} \cite{paolo-digital-twin-cloud-edge-serverless} introduces a microservices-based and Serverless-ready model for \textit{DTs}, establishing the foundation for cost-effective \textit{DT} deployment. The research results show that the combined use of microservices and serverless computing has significant potential to address challenges such as accommodating variable application requirements. The authors implemented a hypothetical Digital Twin setup combining two Serverless deployments, one relying on a cloud provider Serverless function and one using a Serverless framework at the edge, and one micro-service-based deployment. The findings show that the Serverless implementation performs an order of magnitude worse than microservices. The proposed study lacks a real Digital Twin use case implementation, does not provide any serverless abstraction for \textit{DT} data model definitions, and does not discuss event storage.

{\color{black}

The Digital Twin literature still has many unexplored study areas, missing an actual deployment of a Digital Twin implementation built on top of Cloud-Native solutions. To the best of our knowledge, no prior work has proposed an event-driven, Serverless-based architecture for \textit{DT} systems that provides Serverless abstractions to enable end-users to seamlessly combine flexible model design with system-specific configurations. This work addresses these gaps by introducing a novel solution that includes (1) an extended \textit{DTDL} for higher-level abstraction, enabling end users to define domain entities while maintaining control over underlying services, (2) an automated process to orchestrate \textit{DT} components creation, modification and deletion, including messaging, routing, and data storage, and (3) the development and deploiyment of a Smart City \textit{DT} use case to validate the platform. This novel approach is expected to significantly enhance the adoption of \textit{DT} technology by simplifying the development and deployment of new and diverse use cases.

}

%% file: 3-proposal.tex
\section{KTWIN Architecture Design}
\label{section:ktwin-architecture-requirements-design}

\textit{KTWIN}, a Kubernetes-based Serverless platform for Digital Twins, automates the orchestration of Digital Twin services allowing operators to easily define twin's data models and their relationships reducing the complexity of deploying and managing \textit{DT} systems. {\color{black}The following subsections outline the architectural requirements and decisions of \textit{KTWIN} and detail its main components. Additionally, they describe the implemented custom resources, event routing mechanisms, and key aspects of data management and storage.}

\subsection{Architectural Requirements and Design Choices}
\label{section:architectural-definitions}


{\color{black}

Understanding the distinct needs of the personas involved in the design, development, and operational lifecycle of Digital Twins is instrumental in shaping \textit{KTWIN}’s architectural requirements and in ensuring that the platform is tailored to meet their specific challenges and objectives.
Therefore, we defined \textit{KTWIN} revolving around three primary personas, described as follows: 
\begin{itemize}
    \item The \textit{DT Domain Expert} is primarily responsible for the strategic direction and governance of the Digital Twin ecosystem. Their responsibilities include defining the functional requirements by modeling entities and describing their behaviors, often leveraging an ontology-based language to ensure precise representation of domain-specific entities. This requires a system capable of adapting to diverse domains through open and flexible modeling tools. 
    \item The \textit{DT Developer} focuses on implementing and validating services that encapsulate the domain logic defined by the \textit{DT Domain Expert}. The developer seeks for abstractions such as Function as a Service (\textit{FaaS}), service containerization, and the usage of Software Development Kits (\textit{SDKs}), to streamline service creation and deployment.
    \item Lastly, the \textit{DT Operator} responsibilities include the deployment and maintenance of the Digital Twin solution, combining domain requirements with system-level configurations. Their role requires an unified, vendor-neutral specification language that enables seamless deployment across cloud, on-premise, and edge environments. 
\end{itemize}


\subsubsection{Architectural Requirements}

\textbf{Orchestration Abstraction.} The orchestration abstraction provided by \textit{KTWIN} significantly streamlines the process of designing and managing Digital Twin systems by reducing complexity in implementation and maintenance. High-level data modeling, powered by ontology-based and open-standard tools, allows users to define and manage entities, relationships, and behaviors in a structured and reusable manner. This approach ensures cross-domain compatibility, enabling a consistent representation of Digital Twin systems. Furthermore, the simplified configuration process enables users to have an enhanced control and flexibility over the underlying services, combining domain-specific configurations with system-specific requirements to set up and operate the platform effectively. By abstracting underlying complexities, \textit{KTWIN} democratizes access to sophisticated Digital Twin capabilities, reducing barriers to adoption and minimizing implementation and maintenance costs.


\textbf{Self-Contained Modules}. The modularity inherent in the platform ensures that the developers define twin behaviors as self-contained, interchangeable functions, allowing development, deployment, updates, and scaling of the individual services. By defining twin's implementation as independent and self-contained functions, \textit{KTWIN} enhances maintainability, as updates or replacements to a module can be performed with minimal disruption. Additionally, modularity fosters reusability, allowing components to be repurposed across multiple Digital Twin implementations, reducing redundancy and accelerating development cycles.


\textbf{Data Acquisition and Ingestion}. Data acquisition and ingestion in \textit{KTWIN} are critical for integrating real-world data into Digital Twin systems, ensuring accurate and up-to-date models. The platform supports diverse data sources, including IoT devices and edge systems, through a centralized broker that supports multiple protocols such as MQTT, HTTP, and AMQP. This capability allows \textit{KTWIN} to collect, route, and process events efficiently, whether they originate from high-frequency sensors or intermittent data sources. Furthermore, the system can adapt to increasing data volumes and evolving workload demands effectively, allowing individual components to scale independently based on real-time demand. The event-driven design also ensures resilience, as services can continue functioning even during high loads or partial failures.


\textbf{Services Interoperability}. Interoperability enables seamless integration and operation across diverse systems and environments. By adhering to open standards and leveraging cloud-native tools, the platform ensures compatibility with a broad range of technologies, promoting portability and reducing vendor lock-in. Additionally, \textit{KTWIN} supports deployment across multi-cloud, edge, and on-premises infrastructures, allowing organizations to tailor the platform to their specific operational needs. This capability ensures that Digital Twin systems can operate efficiently regardless of the underlying environment, making \textit{KTWIN} a versatile solution for heterogeneous ecosystems.

The aforementioned requirements provide a clear understanding of the key architectural considerations that shaped the design and development of \textit{KTWIN} platform. They played a pivotal role in guiding the architectural decision-making process, detailed as follows. 

\subsubsection{Architectural Design Choices}

\textbf{Serverless}. \textit{KTWIN}-based applications requires high scalability, low operational overhead, and the ability to handle variable workloads efficiently. Serverless computing platforms provide Function(s)-as-a-Service (\textit{FaaS}) by hosting individual callable functions. These
platforms provide reduced hosting costs, fast deployment, automatic scaling, high availability, fault
tolerance, and dynamic elasticity through automatic provisioning and management of compute infrastructure \cite{serverless-2018}, making it ideal for handling variable workloads with low operational overhead. The proposed \textit{KTWIN} Serverless architecture implements automated services infrastructure provisioning based on the provided ontology resources definition. The usage of vendor-specific Serverless platforms come with certain limitations, including cold start latency, execution time restrictions, and the risk of vendor lock-in \cite{serverless-2018}.

\textbf{Event-Driven}. Event-Driven Architecture (\textit{EDA}) allows to decouple services, scale effectively, and process events in real-time. \textit{KTWIN} employs a centralized broker ensuring the processing of events regardless of their source and supporting multiple communications protocols, including MQTT, HTTP and AMQP. Additionally, its event-based routing mechanism is automatically configured based on ontology's relationship definitions, simplifying development and operation efforts for the platform users. The \textit{KTWIN} event-driven architecture provides the flexibility and resilience necessary for Digital Twin systems, allowing independent service evolution, fault tolerance, and improved extensibility. Furthermore, the event-driven microservices design seamlessly integrates with Serverless architectures, offering inherent benefits such as high availability, fault resilience, and cost efficiency.

%

\textbf{Kubernetes and Container-based Services}. Kubernetes serves as a backbone for \textit{KTWIN}, providing scalability, flexibility, and resilience to manage complex containerized applications across diverse environments. \textit{KTWIN} leverages on Kubernetes operators and Custom Resource Definitions (\textit{CRDs}) to automate the management and provisioning of cluster infrastructure. Kubernetes also enables multi-cloud and edge-based deployments strategies needed for microservice-based architecture while avoiding vendor lock-in. Lastly, Kubernetes integrates with rich cloud-native environments, including Serverless platforms. This integration enables \textit{KTWIN} to leverage Function-as-a-Service (FaaS) models.

\textbf{Cloud Native-based Open Platform}. \textit{KTWIN} leverages the rich Cloud Native ecosystem to meet scalability, flexibility, and resilience needs. The adoption of open solutions and standards benefits from an open, innovative and vendor-neutral ecosystem that encourages innovation. The rich open ecosystem also provides cost-efficient and mature tools to manage deployment, observability and security. In terms of ontology-based domain definition languages, \textit{KTWIN} leverages the Digital Twins Definition Language (\textit{DTDL}) to offer a flexible and open approach for describing digital twin models of smart devices. \\


\textit{KTWIN} advances the state-of-the-art on \textit{DT} platforms by addressing key limitations in existing approaches while introducing innovative capabilities that have yet to be thoroughly explored in the literature. It achieves this by (1) reducing operational costs through the automation of manual processes, streamlining deployment and management tasks, (2) offering enhanced control and flexibility over the container-based services that constitute the \textit{DT} landscape, enabling tailored configurations to meet diverse requirements for DTs of any domain, and (3) supporting deployment across any Kubernetes cluster, whether on-premises, at the edge, or in the cloud, minimizing the risk of vendor lock-in, and providing organizations with greater autonomy and adaptability.

}
\subsection{KTWIN Main Components}

\begin{figure}[ht]
    \centering
    \includegraphics[scale=0.35]{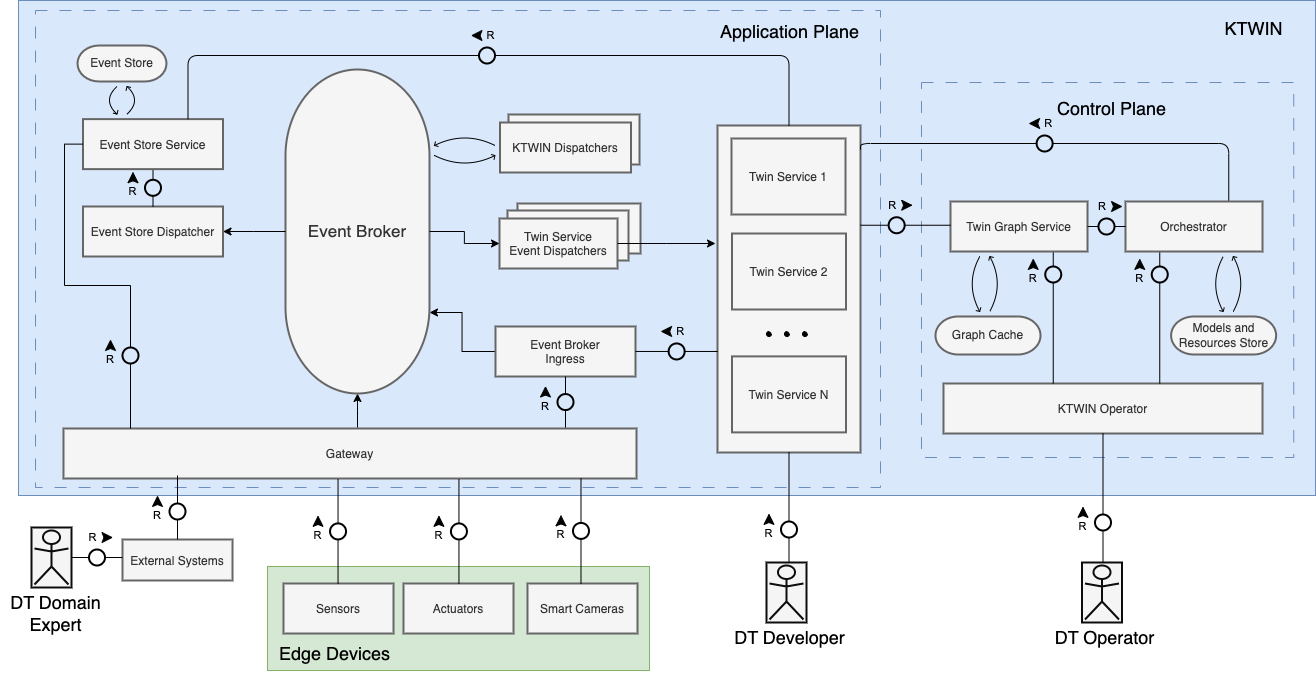}
    \caption{Proposed high-level KTWIN Architecture.}
    \label{fig:ktwin-architecture}
\end{figure}

This section explains the high-level architecture definition proposed for \textit{KTWIN}, which can be divided into Control Plane and Application plane components as depicted in Figure \ref{fig:ktwin-architecture}. The Control Plane components allow human operators to define the Digital Twin entities and their relationships through a generic resource definition file, allowing the deployment of \textit{DTs} of any domain knowledge. The human operator interacts with the \textit{KTWIN Operator} which is responsible for orchestrating the creation of all the underlying components and services from the Application Plane required to deploy the Digital Twin scenario. The operator contains the \textit{KTWIN}-related resources knowledge, meaning that it knows all the steps and underlying resources required to instantiate a new twin within the platform. The Operator executes creation, modification and deletion requests against the \textit{Orchestrator} component.

The \textit{Orchestrator} applies and maintains the desired resource states defined by the operator which incurs creating a new \textit{Twin Service}, maintaining the expected number of Event Dispatchers, or keeping the Event Broker routing rules up-to-date. The configured resources are persisted into the \textit{Orchestrator} storage. In addition, the definition of the resource model includes domain-related entities, their corresponding data model attributes, the different instances of an entity, and their relationships with other entities. The entity's relationships compose a Twin Graph. Because Application Plane components require the Twin Graph information to process and propagate events to related entities, the graph data is replicated to a fast access cache storage. Hence, Application Plane components can consume the graph information through the \textit{Twin Graph Service}, decoupling the control plane storage from the application plane components and offering higher scalability and faster response time.

The Application Plane components are responsible for processing the Digital Twin events and storing them for further data analysis and visualization. Real-world edge devices interact with \textit{KTWIN} by sending events that represent a state change in some defined entities of the Digital Twin. Devices such as an actuator can also subscribe and receive events from \textit{KTWIN} indicating that some action must be executed in the edge. The events are published to or subscribed from the \textit{Event Broker} through the \textit{Gateway} component. The \textit{Gateway} acts as a load balancer from ingress requests and redirects the traffic according to the protocol or endpoint: while HTTPS requests are redirected to the \textit{Event Broker Ingress} which publishes them to the \textit{Event Broker}, AMQP and MQTT connections are directly forwarded to the \textit{Event Broker}. Incoming broker requests are routed to the corresponding target based on in-broker routing rules that send events to different in-broker queues and are subscribed by \textit{Event Dispatchers}. The \textit{Event Dispatchers} are lightweight components responsible for subscribing events and dispatching them to the corresponding \textit{Twin Service} service deployed by the end user.

The \textit{Twin Service} are designed and deployed as a \textit{Serverless} function that contains a custom implementation of some Digital Twin domain defined by the \textit{DT} domain expert. Each \textit{Twin Service} is associated with a real twin and is responsible for processing one more event type. \textit{Twin Services} can communicate with its real-world counterpart or with another \textit{Twin Service} in which it has a relationship. In addition, they can be scaled down to zero or up to thousands of instances, offering improved performance and efficiency in resource usage.

The \textit{Event Store} is a highly scalable data store designed to save the large number of events generated by real-world devices and processed by Virtual Services. Events persisted to the \textit{Event Store} are queued with the broker and subscribed by the \textit{Event Store Dispatcher} that dispatches them to the \textit{Event Store Service}. In addition, end-users can implement \textit{Twin Service} services to obtain the latest status of some twins using \textit{Event Store Service} APIs. The \textit{Event Store} APIs also allow external systems to fetch persisted events in batches.

\textit{KTWIN} provides support for clients to publish events using a variety of protocols, such as MQTT, AMQP, and HTTP. In this context, the \textit{KTWIN Dispatchers} are event dispatchers built to implement protocol conversion. For instance, they convert real-world MQTT events published by devices into in-cluster events, as well as in-cluster events back to MQTT device events. They are required because \textit{KTWIN} uses a set of event header metadata information to route messages to the corresponding target service, which cannot be implemented using the MQTT4 protocol. In contrast, the device's AMQP-generated events can be directly published into \textit{Event Broker} since AMQP offers routing headers support. Lastly, the \textit{Event Broker Ingress} is a request ingress service used by \textit{Twin Services} or real devices to publish events into the \textit{Event Broker}, not requiring them to establish a stateful connection with it, for instance, HTTP. More information about these components and their implementation is provided in the next section.

\subsection{Digital Twin Resources Definition and Event Routing}

\textit{KTWIN} provides resources and data abstractions for Digital Twin operators to define their own Digital Twin models and relationships allowing the deployment of any \textit{DT} domain. It uses the open-source Digital Twin Definition Language (\textit{DTDL}) as the base entity definition framework and implements twin message routing rules on top of these definitions. In addition, \textit{KTWIN} enhances the existing specification language by adding some system's related information that can be configured to customize the deployment settings of a Digital Twin use case.

The \textit{KTWIN} resource definition framework is divided into two main definition resources: \textit{Twin Interface} and \textit{Twin Instance}. The \textit{Twin Interface} describes the attributes, relationships, and other types of contents common for any Digital Twin of that type, being reusable. A \textit{Twin Instance} is an instance of some interface in the real world. In this context, a \textit{Twin Interface} Car has attributes, such as color, year, and length, and relationships, such as owner and wheels. A \textit{Twin Interface} can be created for each car that belongs to a Digital Twin, where each vehicle is unique and has its corresponding representation in the real world. In addition, users may also define system-specific settings such as allocated resources and auto-scaling attributes.

The \textit{Twin Interface} definition allows domain experts to describe properties, relationships, commands, and inherit a parent interface. The \textit{Twin Instance} definitions allow users to create Instances of interfaces, and configure relationships with other instances. Tables \ref{table:twin-interface-specification-table} and \ref{table:twin-instance-specification-table} show the fields in each resource specification.

\begin{table}[ht]
\centering
\caption{Twin Interface Specification Fields}
\begin{tabular}{|p{4cm}|p{9cm}|}
 \hline
 \textbf{Field} & \textbf{Description} \\ [1.2ex]
 \hline
   Name & The Twin Interface Name unique identifier \\
 \hline
   Properties & List of properties that define the Twin Interface. A property has a name, description, and data type or schema definition (in case of an enumeration). \\
 \hline
   Relationships & List of relationships with other Twin Interfaces. A relationship has a name, a description, a target Twin Interface, and a multiplicity indicator. \\
 \hline
   Commands & List of commands and actions that can be executed in a Twin Interface. Only Twin Interfaces with a common relationship can trigger a command of a Twin Interface. A Command has a name, description and schema definition. \\
 \hline
   Parent Twin Interface & It is possible for a Twin Interface to extend another Interface. In this context, all properties, relationships, and commands are inherited by the child Interface. \\
 \hline
   Service Settings & It is possible for operators to define the container implementation that is going to be used to execute events of the specific Twin Interface, CPU and memory reserved for the container, as well as auto-scaling metrics. \\
 \hline
   Routing and Persistence Settings & Operators may also define whether a specific type of event must be persisted in the Event Store, or if some Instance's events must be propagated to an Instance with a relationship for data aggregation.  \\
 \hline
\end{tabular}
 \label{table:twin-interface-specification-table}
\end{table}

\begin{table}[!]
\centering
\caption{Twin Instance Specification Fields}
\begin{tabular}{|p{4cm}|p{9cm}|}
 \hline
 \textbf{Field} & \textbf{Description} \\ [1.2ex]
 \hline
   Name & The Twin Instance Name unique identifier. \\
 \hline
   Twin Interface & The Twin Interface that the Instance is associated with. \\
 \hline
   Properties & The list of properties defined in the Twin Interface. The Twin Instance property has the same name as defined in the Instance and its corresponding value. \\
 \hline
   Relationships & The list of relationships defined in the Twin Interface. A relationship has a name, the target Twin Interface type and the target Twin Instances. \\
 \hline
\end{tabular}
 \label{table:twin-instance-specification-table}
\end{table}

The resources described previously are used by \textit{KTWIN Operator} to build an event routing mechanism without any manual intervention from the \textit{DT} operator. The messaging routing rules support the following communication flows: (1) an event generated by the real device instance and sent to the \textit{Twin Service} service, (2) an event generated by the \textit{Twin Service} instance and sent to the corresponding real device, (3) an event generated by some \textit{Twin Service} and sent to another \textit{Twin Service} (these twins must have some relationship), and (4) an event generated by some \textit{Twin Service} and published to Event Store. \textit{KTWIN} uses the \textit{Twin Interface} and \textit{Twin Instance} definitions as well their relationships to build event propagating and routing built-in rules within the \textit{Event Broker}.

\textit{KTWIN} divides events into four subcategories which identify the context in which an event happened: real, virtual, command, and store. The subcategories information is included in the event message, together with the \textit{Twin Interface} and the \textit{Twin Instance} identifiers to identify the source of the event and determine its targets. In case of an MQTT connection, real devices publish events to the \textit{Event Broker} topic containing the \textit{Twin Interface} and the \textit{Twin Instance} information the event is associated to, such as \textit{ktwin.real.<twin-interface>.<twin-instance>}. For HTTP and AMQP requests, this information is provided in the message headers. Events published by devices following the previously mentioned pattern are routed to the corresponding \textit{Twin Service}. The same devices can subscribe to events generated by its Twin Service counterpart in the topic \textit{ktwin.virtual.<twin-interface>.<twin-instance>}. 

\begin{figure}[ht]
    \centering
    \includegraphics[scale=0.35]{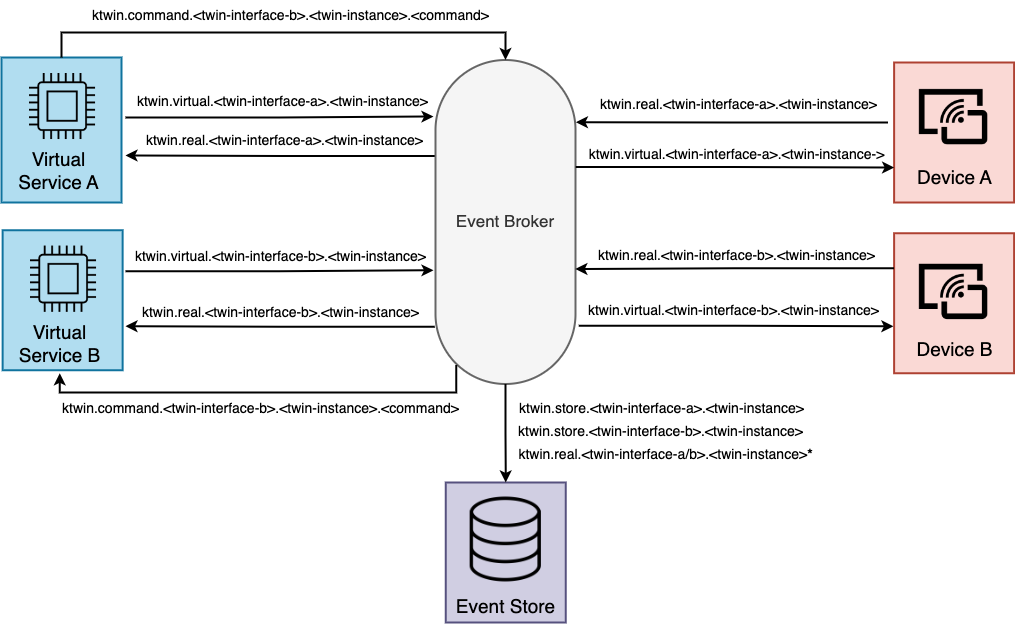}
    \caption{KTWIN Message Routing implemented in Event Broker.}
    \label{fig:ktwin-message-routing}
\end{figure}

\textit{Twin Service} with a common relationship in the twin graph can propagate events to each other using the command event category, with the following pattern in the message header: \textit{ktwin.command.<twin-interface>.<twin-instance>.<command-name>}. In addition, events that are supposed to be routed to the event store for persistence use the following routing header: \textit{ktwin.store.<twin-interface>.<twin-instance>}. Finally, in case some \textit{Twin Interface} does not require computing any information with the generated device event, users may configure \textit{KTWIN} to route real events directly to the \textit{Event Store} using the routing and persisting settings in the \textit{Twin Interface} specification. Figure \ref{fig:ktwin-message-routing} shows how the routing implementation works with hypothetical examples of A and B entities.

Although the routing keys contain information about the \textit{Twin Instance} and the \textit{Twin Interface}, the in-broker routing is mostly done at the interface level, thus reducing the number of routing rules required within the broker. In this context, the routing is implemented using a start-with logic, such as \textit{ktwin.<event-category>.<twin-interface>.*}. Once the user-implemented service receives the event, the routing key is decoded from the message, so the application logic knows the instance owner of that message. The application later uses this information for data propagation to the corresponding related entities using command events or data persistence in the event store. In this context, in case some command event must be generated, the application must have access to the graph relations. The previously described steps introduce some complexity for end-users to implement this logic, hence \textit{KTWIN} offers a Software Development Kit (\textit{SDK}) to be imported by end-users when writing their functions. The \textit{SDK} implements the routing key's decoding and command-based event propagation based on twin graph data in a user-friendly way.

\subsection{Data Management and Storage}

The Digital Twin Definition graph defined by the domain experts is stored within the Control Plane store and maintained by \textit{Orchestrator}. It allows \textit{DT} operators to easily manage and visualize the defined \textit{Twin Interfaces}, their corresponding instances, attributes and commands. However, the Twin Graph information must also be available for the application plane services implementations during the event processing. Hence, the Digital Twin graph data is replicated and cached in the \textit{Twin Graph Cache} to avoid impacts of Application Plane incoming requests to Control Plane components, reducing the response time and minimizing the control plane disruption. The graph cache implementation uses a key-value in-memory data store.

Application Plane twin events are persisted into a scalable and reliable \textit{Event Store} database. Events are stored in a time series format containing the history of all events generated for a specific instance during its lifespan. Events are asynchronously published to \textit{Event Broker} and routed to the \textit{Event Store Dispatcher} before being persisted in the database by the \textit{Event Store Service}. In addition to the event payload, the persisted data includes the timestamp in which the event occurred, and the \textit{Twin Instance} and \textit{Twin Interface} that the events belong to.

In addition to data persistence, the \textit{Event Store Service} provides APIs to consume past events persisted in the data store. User-defined services can query the latest status of some \textit{Twin Instance} stored and perform business logic using this data while processing incoming events. Additionally, external applications can consume a batch history of events of a specific \textit{Twin Instance} or query by events of several instances of the same interface for further data analytics and machine learning. Data analytics and machine learning discussions are out of the scope of this initial research and are future research opportunities for \textit{KTWIN} use cases.

%% file: 4-implementation.tex
\section{KTWIN System Design and Implementation}
\label{section:ktwin-implementation}

This section describes the high-level \textit{KTWIN} prototype implemented in this study. First, it presents the implementation architecture, how each component was implemented, and the corresponding chosen technologies. Next, we show how \textit{KTWIN} offers Digital Twin modeling abstractions using Kubernetes Custom Resource and implements event definitions and routing using Cloud Event specification. The section also presents how collected device data is stored within the \textit{Event Store} to allow further data analysis. Finally, we conclude the section by presenting how \textit{KTWIN} implements observability requirements. The implemented prototype and deployment setup are available at \url{https://github.com/Open-Digital-Twin}.

\subsection{KTWIN Architecture Implementation}

\begin{figure}[ht]
    \centering
    \includegraphics[scale=0.23]{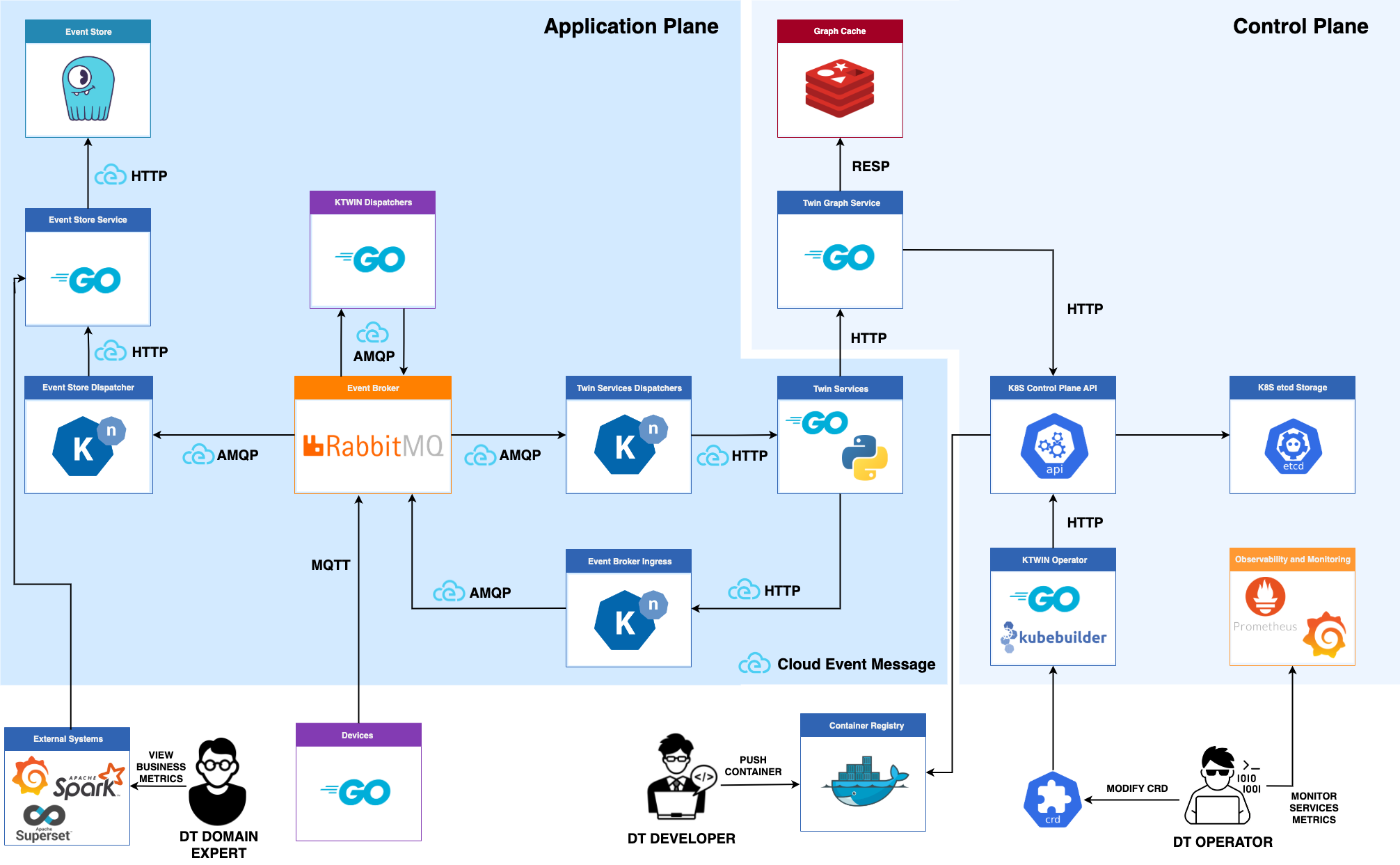}
    \caption{KTWIN Implementation Diagram.}
    \label{fig:ktwin-implementation-diagram}
\end{figure}

The \textit{KTWIN} implemented prototype embraces Cloud-Native technologies allowing end-user applications to be implemented, deployed, and orchestrated as containerized microservices providing resilience and scalability. Figure \ref{fig:ktwin-implementation-diagram} shows the building blocks of the implementation evaluated in this investigation.

The main personas involved in \textit{KTWIN} development and operational process are: \textit{DT Domain Expert}, \textit{DT Developer} and \textit{DT Operator}. The \textit{DT Operator} creates the Custom Resource Definition (\textit{CRD}) files to model the Digital Twin use case within \textit{KTWIN} and apply the resources using Kubernetes CLI. The CRD files are based on the ontology definition originally provided by the \textit{DT Domain Expert}. The provided files are processed by \textit{KTWIN} Operator, developed in Golang using Kubebuilder, an open-source SDK library that facilitates seamless interaction with Kubernetes Control Plane APIs. These \textit{CRD} files are stored in Kubernetes \textit{etdc} storage, ensuring reliable and consistent data management. The \textit{KTWIN} Operator creates the required Application Plane resources to instantiate the defined use case. The container-based functions implemented by the \textit{DT Developer} are pulled from a container registry and deployed as a \textit{Twin Service}. Finally, the \textit{DT Domain Expert} can analyze the processed data using some visualization tool that fetches the data from the \textit{Event Store}.

The \textit{Twin Graph Service} pulls the defined graph stored in \textit{etdc} storage through Kubernetes Control plane APIs, split into subgraphs of \textit{Twin Interfaces} that contain all \textit{Twin Instances} and their relationship and caches in Redis, an in-memory key-value storage. The cache service exposes the subgraph data via Restful API so user-defined services can fetch the graph information of a particular \textit{Twin Interface} while the service starts up. The subgraph content must be immediately available and provide fast data access to reduce impacts on the cold start time of the service.


\textit{KTWIN} Application Plan components are implemented on top of \textit{Knative} \cite{knative-site}, an open-source Kubernetes-based platform designed to build, deploy, and manage modern serverless workloads. \textit{Knative} extends Kubernetes to provide a set of middleware components that enable the deployment and management of serverless applications. It is the core foundation of Cloud Run in Google Cloud Platform (GCP) bringing together the best of both Serverless and containers, such as increasing the developer’s productivity by three times \cite{knative-cloud-run}.

The \textit{Event Broker}, one of the core components of \textit{KTWIN}, was implemented by instantiating a \textit{Knative} Broker resource, which deploys a RabbitMQ broker and provides an endpoint for event ingress to which producers can post their events. RabbitMQ is a scalable and reliable messaging broker \cite{pubsub-iot-rabbitmq} able to handle millions of concurrent IoT connections \cite{rabbitmq-native-mqtt}. In addition, RabbitMQ supports open standard protocols, including AMQP and MQTT, and provides MQTT-AMQP protocol interoperability \cite{rabbitmq-native-mqtt} and header-based routing.

The \textit{Twin Services} were implemented as independent application containers and deployed using the \textit{Knative} Service custom resource. The \textit{CRD} creates the required Kubernetes resources to allow scaling up and down multiple replicas of the container and exposes an HTTP endpoint within the cluster. The \textit{Twin Service} event subscription was implemented using a \textit{Knative} Trigger custom resource. A Trigger allows users to define to each service an event that will be forwarded based on some message metadata. The routing rules are implemented within the \textit{Event Broker} using the RabbitMQ Topology Operator and use Cloud Event \cite{cloud-event-spec-site} specification metadata fields in the routing rules. \textit{Knative} Dispatchers subscribe to events in the Event Broker queues and post them to the corresponding \textit{Twin Services} based on the provided routing rules. \textit{Twin Service} generated events are published to the created event Event Broker Ingress and routed to the corresponding subscriber.

The \textit{Event Store Dispatcher} is a standard subscriber created during the \textit{KTWIN} installation steps that subscribes to events with \textit{Event Store} as the target. These events are posted to the \textit{Event Store Service} and persisted in a ScyllaDB cluster. ScyllaDB is a NoSQL distributed database, compatible with Apache Cassandra, designed for data-intensive applications requiring high performance and low latency \cite{benchmarking-mongodb-scylladb}. Scylla database was deployed using the official ScyllaDB Operator \cite{scylladb-operator} to automate the NoSQL cluster deployment process and operational tasks such as scaling, auto-healing, rolling configuration changes, and version upgrades.

The implemented prototype also includes a set of MQTT publisher's applications developed in Golang and deployed as containers to produce IoT-generate data used during the evaluation scenarios. The containerized applications allow the creation of a configurable number of threads allowing the execution of thousands of MQTT publisher devices within a single container without exceeding Kubernetes pods limits per node. Although \textit{KTWIN} allows devices to publish messages in different protocols, such as HTTP Cloud Event-based messages or AMQP, the implemented devices use MQTT because it is widely adopted in IoT systems due to its lightweight design \cite{iot-data-multi-broker-environments}.

The devices publish MQTT events to RabbitMQ Event Broker which provides AMQP-MQTT protocol interoperability. However, RabbitMQ does not support MQTT routing based on Cloud Event specification headers. To address this limitation, \textit{KTWIN Dispatchers} convert the MQTT message routing, originally based on RabbitMQ routing-key routing, to a header key routing rule. This conversion ensures that MQTT edge device messages are properly converted to Cloud Events. The same limitation does not exist for AMQP or HTTP messages, since they are compatible with Cloud Event spec. This limitation in RabbitMQ opens up possibilities for future enhancements in the proposed implementation.

Finally, \textit{KTWIN} provides observability capabilities and allows DT Operators to monitor the service's health in Grafana dashboards. The Grafana dashboard built for this prototype displays metrics scraped by Prometheus including event throughput per event type, application CPU and memory consumption, and response time.

\subsection{Kubernetes Custom Resources}

A Kubernetes Custom Resource Definition (\textit{CRD}) is a way to extend the Kubernetes API by defining and managing custom resource types. \textit{KTWIN} makes use of \textit{CRD} to allow Digital Twin Domain Experts to define Digital Twin entities, their attributes and relation, in addition to system-specific definitions, such as allocated CPU or memory and auto-scaling policies. The developed \textit{KTWIN} prototype enhances the open-source Digital Twin Definition language (\textit{DTDL}) by adding system-specific configuration to the ontology-based definition language. The implemented prototype has two \textit{DTDL}-based \textit{CRDs}: \textit{Twin Interface} and \textit{Twin Instance}.

The \textit{Twin Interface} custom resource represents some Digital Twin entity and it describes its name, attributes, relationships, commands, service auto-scaling policies, the related container image that implements the interface behavior and the amount of memory and CPU requested for the container. The \textit{Twin Instance} is an instance of some interface in the real world. The instance custom resource contains the instance name, the corresponding \textit{Twin Interface} identifier, and references to the related \textit{Twin Instances} to which there is a relationship. Both custom resources are defined using the YAML format and can be applied to the cluster using the Kubernetes standard client \textit{kubectl}.

The \textit{KTWIN} Operator was built on top of Kubebuilder which provides an \textit{SDK} to interact with Kubernetes Control Plane APIs. When creating a new \textit{KTWIN} custom resource instance, it triggers the creation of a variety of Kubernetes resources in the Application Plane. These steps are orchestrated by the \textit{KTWIN} operator and Kubernetes Control Plane APIs. The implemented solution uses RabbitMQ Topology Operator that allows the creation of in-broker resources such as Queues, Bindings and exchanges using Kubernetes Custom Resources. The \textit{KTWIN} operator uses these custom resources to instantiate the required broker elements to implement the routing rules based on the provided Digital Twin entities and their relationships and commands.

\subsection{Events Definition and Routing}
\label{section:implementation:events-definitions}

\begin{figure}[ht]
    \centering
    \includegraphics[scale=0.28]{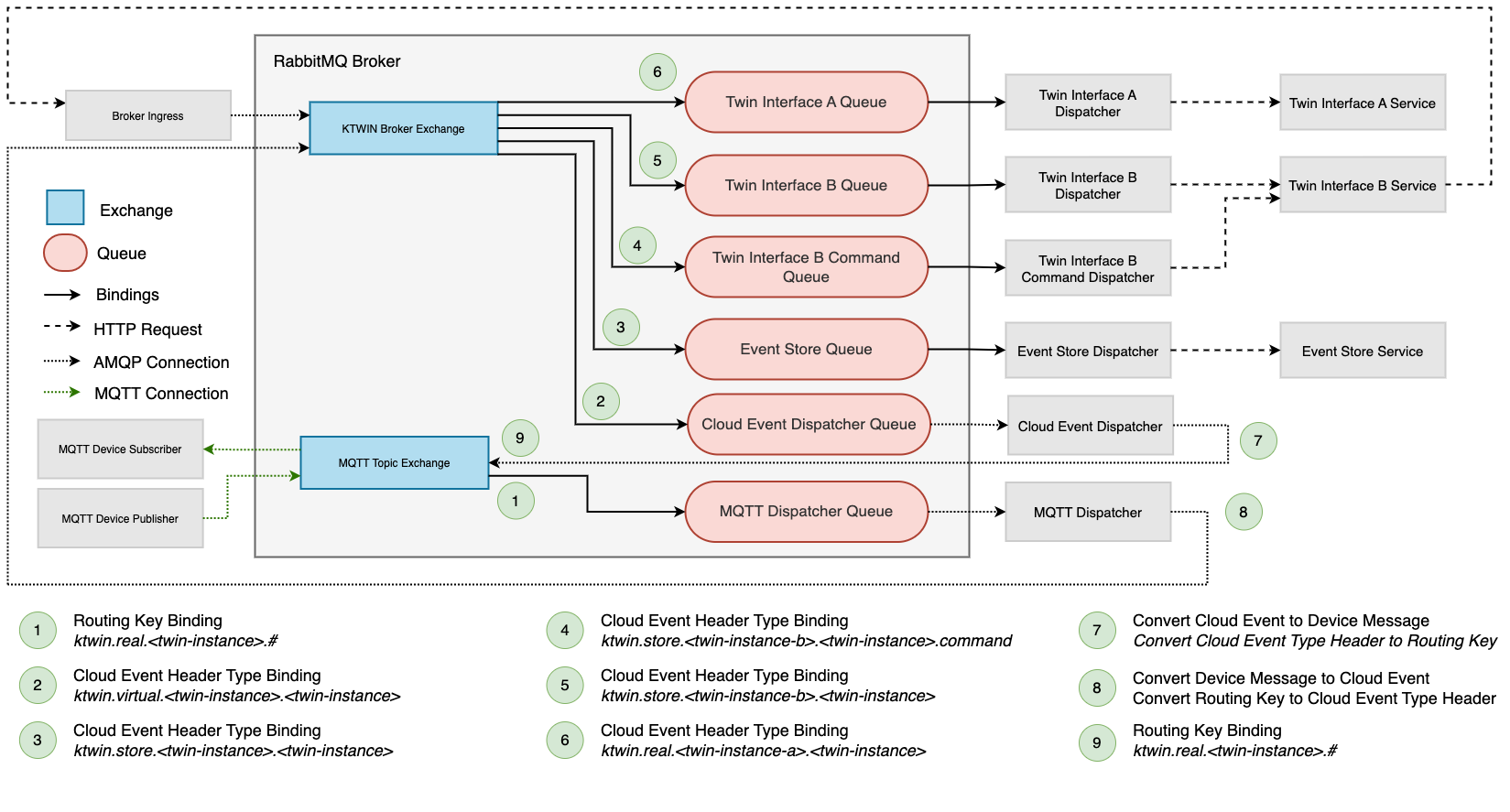}
    \caption{Event Routing Implemented within RabbitMQ}
    \label{fig:implementation:event-routing}
\end{figure}

\textit{KTWIN} events definition and routing rely on in-broker routing mechanisms. Figure \ref{fig:implementation:event-routing} shows how \textit{KTWIN} enhances the routing rules implemented by Knative using RabbitMQ exchanges, queues and bindings. The Broker's internal resources are created using the RabbitMQ Messaging Topology Kubernetes Operator which manages RabbitMQ messaging topologies using Kubernetes Custom Resources.


In Figure \ref{fig:implementation:event-routing}, it is possible to visualize that devices publish events to the MQTT topic \textit{ktwin.real.<twin-interface>.<twin-instance>}. RabbitMQ natively supports MQTT-AMQP interoperability by emulating the MQTT broker via a unique MQTT Topic Exchange, so all publishers and subscribers are virtually connected to the same exchange with a routing key as \textit{ktwin.real.<twin-interface>.<twin-instance>}. \textit{KTWIN} creates routing key-based bindings in the MQTT Topic Exchange to redirect messages to the MQTT Dispatcher Queue for each \textit{Twin Interface} created by the Operator. The \textit{MQTT Dispatcher} subscribes to these routing-key-based messages, converts them to a Cloud Event message, and republishes them to the Broker Exchange. The Broker Exchange uses the Cloud Event \textit{type} header to redirect messages to the corresponding queue and subscribers. The broker exchange bindings include routing rules for (1) events generated by the real instance and redirected by its virtual instance service, (2) events generated by a \textit{Twin Service} and sent to another \textit{Twin Service} (Command), (3) events propagated to the \textit{Event Store}, and (4) event generated by the virtual instance and sent to the real instance. The events generated by \textit{Twin Services} that must be redirected to the real devices are converted from Cloud Event to a routing key message by the Cloud Event Dispatcher, republished to the MQTT Topic exchange, and finally delivered to the device subscribed to the MQTT topic \textit{ktwin.virtual.<twin-interface>.<twin-instance>}.

\textit{KTWIN} provides a Software Development Kit (\textit{SDK}) for handling events and propagating them to edge devices, event stores or related virtual services. The \textit{SDK} implements all the complexity of event handling using the Twin Graph entity relation data, allowing end users to focus on the application logic. \textit{KTWIN} provides an SDK version in both Golang and Python programming languages. An SDK implementation example is provided in the following sections.

\subsection{Graph and Events Storage}

The Twin Graph defined by end-users is stored as CRDs in the Kubernetes \textit{etcd} storage, an open source, distributed, consistent key-value store for shared configuration distributed systems or clusters. The graph data are replicated and cached in the Twin Graph Redis instance, a key-value in-memory cache providing submillisecond response times. Virtual Services require access to the Twin Graph to publish data entities with relationships. This information is fetched from Twin Graph Service RESTful APIs while the service start-up, hence requiring an ultra-fast response time to reduce impacts in the cold start. The Twin Graph information does not need to be fetched by the end-user application code, since KTWIN \textit{SDK} implements this logic and provides a user-friendly interface to consume these data.

The \textit{Event Store} database stores the entire history of \textit{Twin Instances} events during its life cycle. \textit{Event Store} was implemented using ScyllaDB, an open-source NoSQL database for data-intensive apps that require high performance and low latency. ScyllaDB provides high availability, scalability, and efficient resource utilization, being used for a large number of use cases, including real-time analytics and IoT data storage. User-defined services can consume the latest state of some twin event by \textit{Event Store Service}. In addition, external systems can fetch a batch of \textit{Twin Instance} persisted events using the same APIs.

\subsection{Observability Solution}

Observability is crucial for understanding the health, performance, and behavior of distributed systems. \textit{KTWIN} Observability was implemented using Prometheus and Grafana. Prometheus is an open-source monitoring and alerting toolkit designed for reliability and scalability. It scrapes metrics from various services and stores them in a time-series database, enabling real-time monitoring and alerting. Grafana is an open-source analytics and monitoring platform that allows users to define and configure dashboards to visualize Prometheus metrics. Together, Prometheus and Grafana provide a powerful observability stack that allows \textit{KTWIN} users to gain insights into the system, detect issues early, and ensure optimal performance.

\textit{KTWIN} provides a dashboard that allows the \textit{DT} Operator to monitor the number of different event types processed, the CPU and memory usage of the service, as well as its response time. The dashboard was built and used during the evaluation scenarios of this investigation. Prometheus scrapes \textit{KTWIN} components metrics data through the metrics endpoint at regular intervals defined in its configuration. The scraped data is then stored in its time-series database to be consumed by Grafana.

%% file: 5-evaluation.tex
\section{Evaluation Methodology and Results}
\label{section:ktwin-evaluation-results}

In this section, the evaluation methodology of this research is described. The evaluation involves designing, implementing, and deploying a Smart City Digital Twin use case using \textit{KTWIN}. First, this section describes in more detail the ontology of the selected use case as well as the implementation details, followed by the different evaluation scenarios, and the evaluation setup. The final sections present the evaluation results of the executed experiments, including service scalability and the \textit{KTWIN} Serverless capabilities analysis.

\subsection{Smart City Evaluation Use Case}

In order to evaluate the proposed solution, a Smart City use case was designed, implemented, and deployed using \textit{KTWIN}. The selected use case was based on the \textit{DTDL} ontology for Smart Cities developed by Open Agile Smart Cities (\textit{OASC}) and Microsoft \cite{dtdl-smart-cities}. The open-source ontology model was publicly released to accelerate the development of Digital Twin-based solutions for Smart Cities. The \textit{DTDL} is composed of ontology-based definitions from \textit{NGSI-LD} information model specification \cite{ngsi-LD-smart-cities} and \textit{ETSI SAREF} ontologies for Smart Cities (\textit{Saref4City}) \cite{techreport-saref4city}. The final model comprises entities in Urban Mobility, Environment, Waste, Parking, Buildings, Parks, Ports, City Objects, Administrative Areas, and more. A summary of the corresponding group of \textit{Twin Interfaces} created for the evaluation scenarios is presented in more detail as follows.

{\color{black}

\textbf{Administrative Area}: set of entities that represent the organizational structure of a city. At the highest level, the City represents a large human settlement comprising administrative regions such as neighborhoods or districts. A Neighborhood is a geographically localized community within a city, suburb, or rural area, functioning as a smaller administrative unit. Neighborhoods can compute and aggregate data from various entities, such as monitoring air quality in specific regions or tracking the availability of public parking spots in designated areas.

\textbf{Utilities}: this category encompasses a set of entities designed to provide essential resources for residents. A Smart Pole is located on a specific street within a neighborhood and is equipped with various sensors to monitor environmental quality metrics, including air quality, noise levels, crowd density, traffic flow, and weather conditions. Additionally, a Smart Pole can integrate with an Electric Vehicle (\textit{EV}) Charging Station to support sustainable transportation. Streetlights are associated with Smart Poles, enabling the monitoring of lighting levels across different city regions. If a lamp becomes defective, the system can detect it and provide actionable information for timely replacement.

\textbf{Environment}: set of entities used to measure the environment quality observed in certain neighborhood of the city. The Air Quality Observed represents the air quality in a certain region of the city in the determined period. It registers the density of different gases in a certain area such as Carbon dioxide (CO2), Carbon monoxide (CO), and Sulfur dioxide (SO2). The implemented service classifies the Air Quality Index (\textit{AQI}) \cite{airnow-aqi} of the area into Good, Moderate, Unhealthy for Sensitive Groups, Unhealthy, Very Unhealthy, and Hazardous based on the gas density observed by the sensors. The service also updates the air quality index of the neighborhood interface to which it is related. The weather variables observations of a smart pole region are registered in the Weather Observed interface. The interface contains weather-related information such as temperature, precipitation, humidity, snow height, wind direction and speed, and atmospheric pressure. The implemented service aggregates the measured data and computes additional information such as the dew point, feel likes temperature, and pressure tendency. Lastly, the noise level observed entity contains information about the sound pressure level in a certain region of the city.

\textbf{Mobility}: set of entities used for monitoring and managing urban mobility. The Crowd Flow Observed entity tracks crowd congestion, including the number of people, flow direction, and average speed at locations like subway stations or bus terminals. Traffic Flow Observed monitors traffic conditions, providing data such as vehicle spacing and headway times to assess congestion levels. EV Charging Stations are smart facilities linked to Smart Poles, enabling electric vehicles to charge with status updates like available, in use, or out of service, and compatibility with specific vehicle types. Parking Spots represent designated parking areas, with dynamic statuses (free, occupied, or closed) and can be categorized as Off-Street (e.g., garages) or On-Street (e.g., public roads). Parking Groups optionally organize spots by criteria like floor or street.

\textbf{Cross-Domain:} a set of generic entities that can be shared across different domains. A Device can be any kind of sensor, actuator, or camera that generates data for the system. Devices are associated with the majority of the entities previously listed. For instance, a device was involved in order to measure the air quality of a certain area of the city or to report the parking slot availability. The device interface is designed to monitor different operational data from sensors, for example, each device contains its own battery level that can be measured to monitor the device's health.

The open source \textit{DTDL} ontology was converted to \textit{KTWIN} \textit{CRDs} files using \textit{KTWIN} \textit{CLI} implemented as part of the prototype. Each \textit{DTDL} entity was mapped to a \textit{Twin Interface} and a set of \textit{Twin Instances}, and later applied to the cluster to be provisioned by the \textit{KTWIN} Operator. The \textit{CRDs} files mostly include the entity's data model definitions, such as attributes, data-type definitions, and existing interface relationships. The Smart City design effort also included the creation of additional properties and commands, not available in the original \textit{DTDL}, to fulfill the events definition scenarios described in Table \ref{table:evaluation-scenario-events-table}. Finally, a set of container-based applications were developed using \textit{KTWIN} \textit{SDK}, a library used to abstract event handling, propagation and persistence within \textit{KTWIN}. The containers were implemented as stateless functions to handle events for the deployed \textit{Twin Instances}, and defined in the deployment specification files. The same artifacts included parameterized settings to configure auto-scaling policies and control the allocated CPU and memory. The generated artifacts for the prototype evaluation are available on GitHub \footnote{\url{https://github.com/Open-Digital-Twin/ktwin-article}}.
}

\subsection{Evaluation Scenarios and Metrics} \label{section:evaluation-scenarios}

The evaluation scenario of this research aims to reproduce a set of the rich number of events that happen during a day-in-the-life of a Smart City. The simulation design took into account (1) events that are generated throughout the day on a defined frequency, contributing to a fixed data traffic, and (2) events that are generated at a specific moment of the day and do not follow a well-established frequency or are generated because some event criteria in the system were reached, contributing to unpredictable data traffic. In this context, the environment-measured data such as air quality index, city weather, temperature, and noise level observed are events generated at a fixed frequency throughout the full day. However, events directly related to city mobility, such as observed crowds and traffic flows, changes in parking spot availability, and the monitoring of available electric vehicle charging stations, present variable frequencies during the day depending on the city's peak and off-peak hours. Another example of events that present unpredictable traffic is the monitoring of the battery level of sensor devices - each device has its battery level consumption and recharge particularities, so it is difficult to predict when this kind of event will be generated. In-system events, such as the Air Quality Index warning alert when some Smart Pole identifies a low-quality air measurement, are also part of the evaluation scenario.

Table \ref{table:evaluation-scenario-events-table} shows the events implemented and analyzed in the evaluation scenario with their corresponding time interval variation for peak and off-peak hours. The selected event interval considered for simulation compression was from 24 hours to 24 minutes, which means that a 10-second interval represents ten minutes in the real world, allowing the execution of several evaluation experiments in a reduced timeframe. The experiment divides the day simulation into six windows of four hours each, corresponding to 240 seconds in the simulation time. Each event type may differ in frequency within each time window, allowing the reproduction of the variation of events at different moments of the day, such as a city's peak and off-peak hours. In addition to the generated MQTT device events, \textit{Twin Services} generates virtual events back to the edge devices and publishes events for command executions and data persistence.

\begin{table}[ht]
\centering
\caption{Type of Events implemented and analyzed during the Smart City evaluation scenario}
\begin{tabular}{|p{10cm}|p{3cm}|}
 \hline
 \textbf{Event Type} & \textbf{Interval} \\ [1.2ex]
 \hline
   Air Quality observed in the Smart Pole neighborhood & 10s \\
 \hline 
   Noise Level observed in the Smart Pole neighborhood & 10s \\
 \hline 
   Weather observed in the Smart Pole neighborhood & 10s \\
 \hline 
   Crowd Flow observed in the Smart Pole neighborhood & 5s - 30s \\ 
 \hline
   Traffic Flow observed in the Smart Pole neighborhood & 5s - 10s \\ 
 \hline
   Streetlight's on/off status & 720s \\ 
 \hline 
   Electric Vehicle Station availability & 10s - 80s \\ 
 \hline
   Public Parking Spot availability & 5s - 80s \\ 
 \hline
   Update available spots in public parking & 2s - 80s \\ 
 \hline
   Sensor devices battery level health status & 460s \\ 
 \hline
   Air Quality Index warning alert to Neighborhood & In-system event \\ 
 \hline
\end{tabular}
 \label{table:evaluation-scenario-events-table}
\end{table}

Table \ref{table:event-type-experirements} shows all cloud event types implemented and deployed using \textit{KTWIN} in the evaluation scenarios. Each event type contains the \textit{Twin Interface} identifier the event is associated with. In addition, each event has a subtype identifier that is used to route the event to the corresponding target. The subtypes are the following: real, virtual, command, and store. While real devices generate events with real identifiers, virtual events are generated by their virtual counterpart. The event with store subtype indicates that the event must be routed to the \textit{Event Store}, and the command subtype represents some message exchange between two instances with a common relationship. Table \ref{table:event-type-experirements} also contains the throughput range of events per second published to the \textit{Event Broker} for the different city sizes deployed in the evaluation scenarios.

\begin{table}[ht]
\centering
\caption{Cloud Event types implemented in the evaluation scenarios}
\begin{tabular}{|p{8cm}|p{4cm}|p{4cm}|}
 \hline
 \textbf{Event Type} & \textbf{Throughput Range (events per second)} \\ [1.2ex]
 \hline
ktwin.command.city-pole.updateairqualityindex &  5 - 126  \\ 
\hline
ktwin.command.ngsi-ld-city-offstreetparking.updatevehiclecount & 0.5 - 11 \\ 
\hline
ktwin.command.s4city-city-neighborhood.updateairqualityindex &  5 - 118 \\ 
\hline
ktwin.real.ngsi-ld-city-airqualityobserved & 5 - 100 \\ 
\hline
ktwin.real.ngsi-ld-city-crowdflowobserved & 10 - 201 \\ 
\hline
ktwin.real.ngsi-ld-city-device & 7 - 169 \\ 
\hline
ktwin.real.ngsi-ld-city-evchargingstation &  0.2 - 4 \\ 
\hline
ktwin.real.ngsi-ld-city-noiselevelobserved &  5 - 100 \\ 
\hline
ktwin.real.ngsi-ld-city-parkingspot &  0.5 - 10 \\ 
\hline
ktwin.real.ngsi-ld-city-streetlight &  1.7 - 22 \\ 
\hline
ktwin.real.ngsi-ld-city-trafficflowobserved & 10 - 201 \\ 
\hline
ktwin.real.ngsi-ld-city-weatherobserved & 5.5 - 102 \\ 
\hline
ktwin.store.ngsi-ld-city-airqualityobserved &  5 - 125 \\ 
\hline
ktwin.store.ngsi-ld-city-crowdflowobserved & 10 - 208 \\ 
\hline
ktwin.store.ngsi-ld-city-device & 6.7 - 118 \\ 
\hline
ktwin.store.ngsi-ld-city-offstreetparking & 0.5 - 11 \\ 
\hline
ktwin.store.ngsi-ld-city-streetlight & 1.7 - 33 \\ 
\hline
ktwin.store.ngsi-ld-city-trafficflowobserved & 10 - 371 \\ 
\hline
ktwin.store.ngsi-ld-city-weatherobserved & 5 - 127 \\ 
\hline
ktwin.store.s4city-city-neighborhood & 5 - 122 \\ 
\hline
ktwin.virtual.ngsi-ld-city-device & 6.7 - 119 \\
\hline
\end{tabular}
 \label{table:event-type-experirements}
\end{table}

The design of the experiment sizing was based on the New York City dataset of Mobile Telecommunications Franchise Pole Reservation Locations \cite{new-york-mobile-pole-open-data}. The public dataset contains the locations of street light poles, traffic light poles and utility poles reserved by companies authorized by the New York City Department of Information Technology and Telecommunications (\textit{NYCOTI}). The report describes the number of poles installed in each of the five main boroughs: Manhattan, Brooklyn, Queens, Bronx, and Staten Island. Regarding the city mobility metrics, the number of off-street parking spots and electrical vehicle charging stations for the evaluation scenario considered public data maintained by the NYC Department of Transportation (DOT) \cite{new-york-public-parking-facilities}.

In this context, the prototype evaluation of this study used the previously mentioned New York City public data to define the different sizing scenarios for each Twin Interface part of the Smart City Digital Twin Definition model. The evaluation was performed for 1, 5, 10 and 20 neighborhoods to analyze how the system scales when the load increases by 5, 10, and 20 times. In addition, because the experiment timeframe was compressed to a short period of 24 minutes, the number of neighborhoods was reduced to reduce the number of events generated per second at a similar rate. Regarding the amount of generated events, the reduced experiment setup represents approximately a 24-hour city of 1200 neighborhoods. The total number of Twin Instances created during the experiment is presented in Table \ref{table:evaluation-scenario-table}.

\begin{table}[ht]
  \centering
  \renewcommand{\arraystretch}{1.2}
  \caption{Smart City sizing for different numbers of neighborhoods}
  \begin{tabular}{|p{5cm}|c|c|c|c|}
    \hline
    \textbf{Twin Interface} & \multicolumn{4}{c|}{\textbf{\# Twin Instances}} \\
    \hline
    \makecell{Neighborhoods} & 1 & 5 & 10 & 20 \\ \hline
    \makecell{Smart Poles} & 50 & 250 & 500 & 1000 \\ \hline
    \makecell{Streetlights} & 50 & 250 & 500 & 1000 \\ \hline
    \makecell{Air Quality Observed} & 50 & 250 & 500 & 1000 \\ \hline
    \makecell{Crowd Flow Observed} & 50 & 250 & 500 & 1000 \\ \hline
    \makecell{Traffic Flow Observed} & 50 & 250 & 500 & 1000 \\ \hline
    \makecell{Noise Level Observed} & 50 & 250 & 500 & 1000 \\ \hline
    \makecell{Weather Observed} & 50 & 250 & 500 & 1000 \\ \hline
    \makecell{EV Charging Station} & 1 & 5 & 10 & 20 \\ \hline
    \makecell{Off-Street Parking} & 1 & 5 & 10 & 20 \\ \hline
    \makecell{Off-Street Parking Spots} & 20 & 100 & 200 & 400 \\ \hline
    \makecell{Sensor Devices} & 220 & 1110 & 2200 & 4400 \\ \hline
    \makecell{\textbf{Total}} & \textbf{593} & \textbf{2965} & \textbf{5930} & \textbf{11860} \\ \hline
  \end{tabular}
  \label{table:evaluation-scenario-table}
\end{table}

The execution of the experiment was divided into two steps. First, the above scenarios were executed 3 times each with autoscale and scale-to-zero features enabled to verify how the prototype scales for different city sizes using \textit{Knative} Serverless functions. The minimum number of pods per Twin Instances was set to zero and the maximum was set to 18. The \textit{Event Store} service thresholds were set to 1 and 25, because it has a higher throughput. The selected upper thresholds correspond to the maximum amount of pods that could be created considering the memory limits of the nodes set to each they were deployed. The autoscale settings were configured using the concurrency metric, which determines the number of simultaneous requests that each replica of an application can process at any given time, as the value 5. Later, the scenarios of 1 and 5 neighborhoods were executed 3 times each with a fixed number of pods to reproduce scenarios of over-provisioning computing resources (160 pods, 14 per \textit{Twin Instance} + 20 for the \textit{Event Store}) and under-provisioning computing resources (13 pods, 1 per \textit{Twin Instance} + 3 for the \textit{Event Store}). Each simulation aimed to reproduce 24 hours of Smart City events in a 24-minute timeframe.

{\color{black}
The experiments focused on evaluating performance and resource utilization using a set of carefully selected metrics: the number of pods, CPU, memory, and response times. The number of pods metric represents the number of deployed instances in the system and is used to understand how the system scales service instances for different workload sizes. The CPU metrics include CPU requested, reflecting the amount of CPU requested for each service instance, and CPU usage, indicating the actual percentage of resource utilization. Similarly, memory metrics comprised memory requested, representing planned memory allocation, and memory usage, showing real consumption levels. To measure performance under varying loads, response time metrics were analyzed across multiple percentiles: p50 (median) to gauge typical user experience, p90 and p95 to capture performance under higher loads, and p99 to evaluate system behavior during peak or extreme conditions. Together, these metrics provided a comprehensive understanding of the system’s scalability, efficiency, and performance across deployment configurations and neighborhood sizes.
}

\subsection{Evaluation Setup}

The evaluation setup was performed in a private Kubernetes cluster comprising 20 machines. The nodes were divided into three groups based on their sizing and responsibilities: Core (Intel Xeon, AMD Opteron with 24-64+ threads, 16Gb-24Gb RAM), Services (Intel i7/i5 with 4 threads, 4-16Gb RAM) and Devices (Intel i7/i5 with 4 threads, 4-16Gb RAM). Each group of machines runs different workloads and the assignment of the workload pods to the corresponding groups was implemented using Kubernetes node selector based on predefined labels.

The Core group consists of nodes to which the main \textit{KTWIN} components are deployed, such as \textit{Event Broker}, \textit{Event Store}, \textit{Graph Store}, and the \textit{Dispatchers}. The target node is selected during the \textit{KTWIN} installation steps. The Service nodes run end-user services which consist of the \textit{Twin Services} and their corresponding \textit{Event Dispatchers}. For the services, the node selection is implemented by the Operator, so whenever the user defines a new \textit{Twin Interface} resource, the corresponding service is deployed to the target nodes without any user action. Finally, the Device group executes the \textit{IoT} device sensors that publish and subscribe to events of \textit{Event Broker}. By following this approach, it is possible to have more granular control of where each workload will run and based on how much computing power the corresponding node has.

The results of the experiment were collected using Grafana dashboards. The dashboards contain application information, such as the number of requests per second categorized by event types, response time, CPU, and memory usage. The metrics were exposed by the open-source components adopted for the proposed implementation, such as Knative and RabbitMQ, scraped by Prometheus, and consumed by Grafana. Some custom Prometheus metrics were implemented and exposed by \textit{KTWIN} components, such as the number of messages published by \textit{IoT} devices and processed by the \textit{MQTT Dispatcher}.

\subsection{Results and Discussion}

In this section, we present the evaluation results of this research. The results analysis is divided into three sub-sections. In the first sub-section, we present the high-level evaluation results and how each evaluation scenario differs from the others in their sizing, number of connected devices, and processed events per second. Next, we analyze how the system components behaved for scenarios of 1, 5, 10 and, 20 neighborhoods regarding the number of replicas created by the auto-scaler, the amount of CPU and memory requested and used, and service response time. Finally, we discuss the serverless capabilities of \textit{KTWIN} by comparing the previous experiments against under-provisioned and over-provisioned scenarios.

\subsubsection{Evaluation Scenarios Overview}

\begin{figure}[ht]
\centering
\begin{subfigure}{0.42\textwidth}
    \centering
    \includegraphics[scale=0.5]{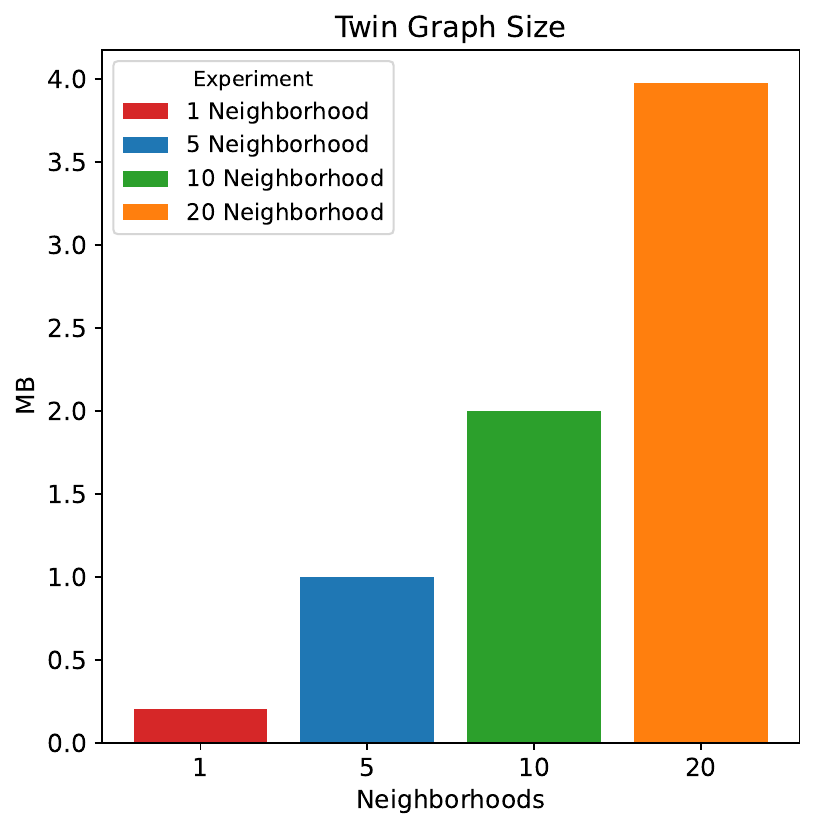}
    \caption{Relationship graph size stored by Graph Cache.}
    \label{fig:evaluation:experiment-1::twin-graph-size}
\end{subfigure}
\begin{subfigure}{0.47\textwidth}
    \centering
    \includegraphics[scale=0.5]{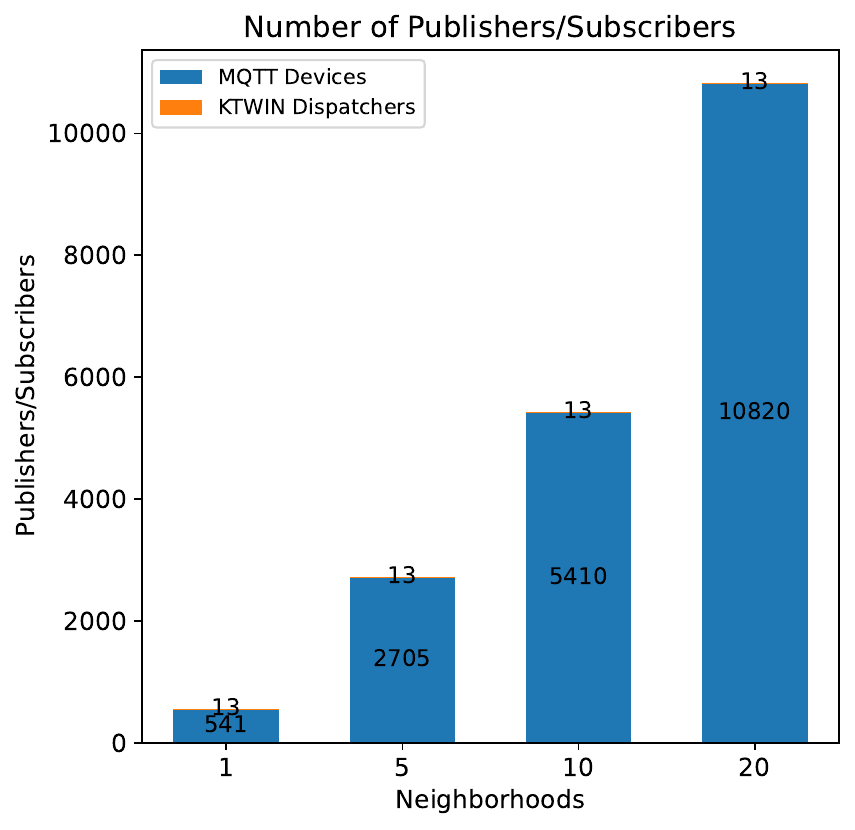}
    \caption{Number of publishers/subscribers connected to the broker.}
    \label{fig:evaluation:experiment-1:number-broker-pubsubs}
\end{subfigure}
\caption{The graph size and publishers and subscribers connected to the broker.}
\label{fig:evaluation:experiment-1-twin-graph-rabbitmq-resources}
\end{figure}

The Digital Twin definitions provided by the domain expert as input for \textit{KTWIN} compose a graph that indicates the relationships of each \textit{Twin Instance}. This information is required to perform data aggregation and propagate events or commands between related instances. The \textit{Twin Graph} data is stored at Kubernetes \textit{etcd} storage and replicated to the \textit{Twin Graph Cache} allowing \textit{Twin Services} to consume the required sub-graph information efficiently. The graph size is proportional to the size of the city. Figure \ref{fig:evaluation:experiment-1::twin-graph-size} shows the Twin Graph size for each experiment in Megabytes. In the 1-neighborhood experiment, the graph takes up 210~kB of storage. The size is multiplied by 5 when compared with the 5-neighborhood experiment representing a size of 1.05~MB. In the scenarios of 10 and 20 neighborhoods, the graphs take a total of 2.10~MB and 4.17~MB respectively, increasing in the same scale of the city size.

Another experiment variable that increases at the same rate as the graph size is the number of \textit{IoT} devices connected to the broker. Figure \ref{fig:evaluation:experiment-1:number-broker-pubsubs} shows the number of publishers and subscribers connected to the broker in each experiment. The chart divides the connected applications into MQTT Devices and \textit{KTWIN} Dispatchers. While the number of MQTT devices increases at the same rate as the experiment's size, the number of dispatchers remains constant. This occurs because \textit{KTWIN} dispatchers process events on the \textit{Twin Interface} level, which means that the same dispatcher processes all events that belong to the same type. The number of \textit{KTWIN} Dispatchers is equal to the number of defined \textit{Twin Interface} plus the 3 standard dispatchers created automatically by \textit{KTWIN} Operator: MQTT Dispatcher, Cloud Event Dispatcher, and Event Store Dispatcher. As presented in the next section, the number of connected devices directly impacts the broker's allocated resources.

\begin{figure}[ht]
\centering
\begin{subfigure}{0.49\textwidth}
    \centering
    \includegraphics[scale=0.4]{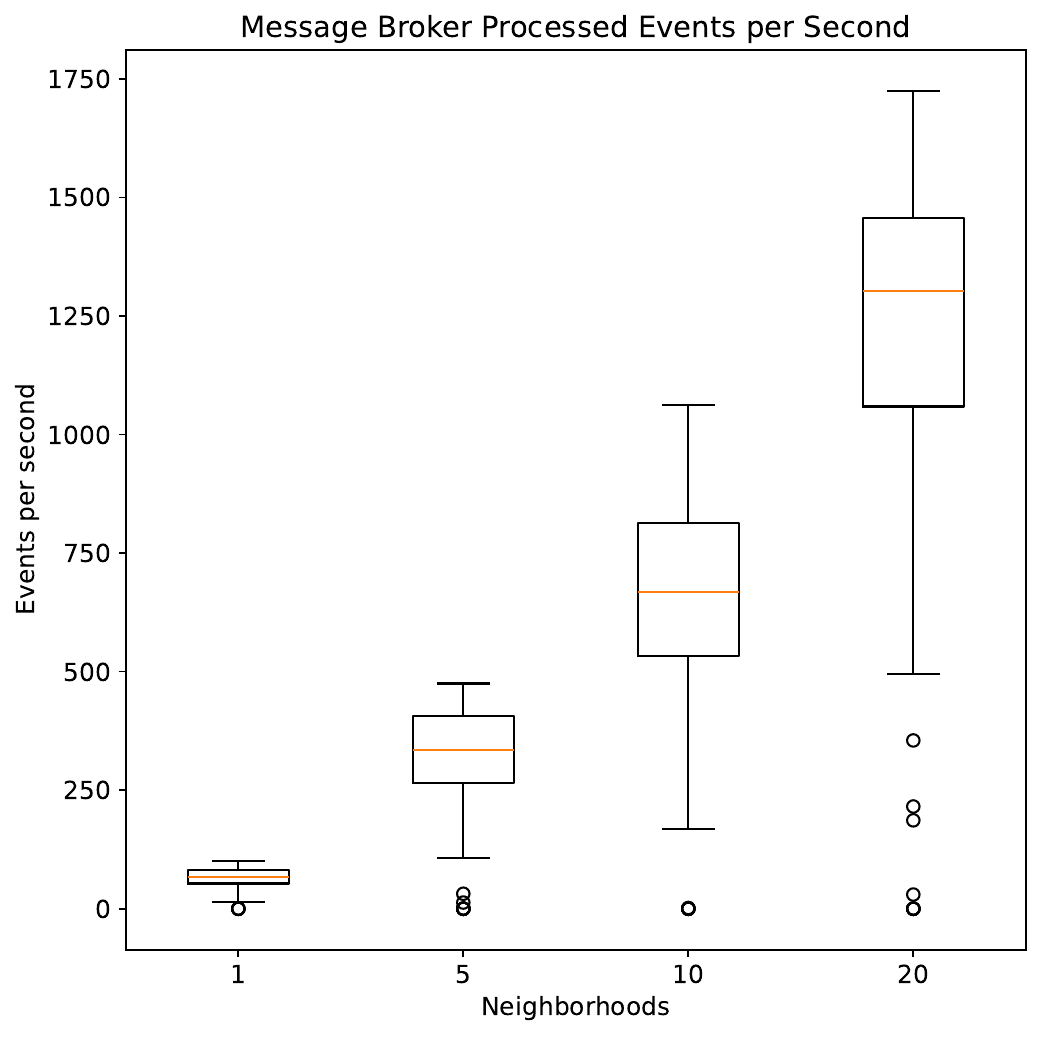}
    \caption{Number of events per second processed by the broker.}
    \label{fig:evaluation:experiment-1:event-per-second-broker}
\end{subfigure}
\begin{subfigure}{0.45\textwidth}
    \centering
    \includegraphics[scale=0.52]{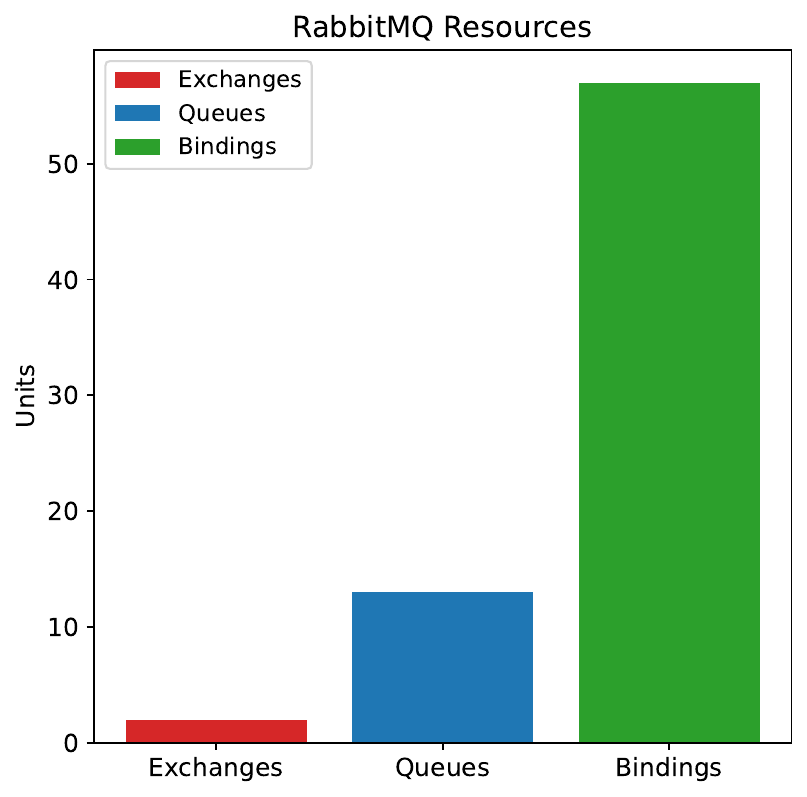}
    \caption{RabbitMQ resources created for each experiment.}
    \label{fig:evaluation:experiment-1:rabbitmq-resources}
\end{subfigure}
\caption{The number of messages processed and RabbitMQ created resources.}
\label{fig:evaluation:experiment-1:pubsub-messagespersecond}
\end{figure}

In the experiments, the devices are responsible for generating events that represent some change in the current status of the real-world twin. As more devices are connected, the higher the probability that the broker receives more events. Figure \ref{fig:evaluation:experiment-1:event-per-second-broker} shows the box plot of the number of events processed per second by the Event Broker. The chart shows that the number of events processed per second in the broker increases approximately in the same ratio of the number of devices connected generating events. 

In the 1-neighborhood scenario, the number of processed events ranges from 13 events per second to 101 events per second, with a median of 67 events per second. In the 5-neighborhood simulation, the median is approximately 5 times the previous scenario, 334 events per second, while the range is between 107 and 475. For the remaining scenarios, the median is 666 and 1301 events per second, while the ranges are between 168 and 1062, and 494 and 1725 events per second. Not all events processed by the broker have devices as sources. In-cluster events to propagate commands, perform data aggregation, and events propagated to the event store are also accounted for in these experiments.

{\color{black}
RabbitMQ performance depends on the number of queues and bindings within the broker. By managing routing rules at the \textit{Twin Interface} level, \textit{KTWIN} ensures that the number of queues scales proportionally to the defined interfaces, demonstrating robust scalability across varying experiment sizes. Figure \ref{fig:evaluation:experiment-1:rabbitmq-resources} illustrates the number of RabbitMQ resources used in all experiments. Regardless of experiment size, the number of RabbitMQ resources remains consistent, minimizing the broker's memory footprint and reducing Kubernetes resource storage. In earlier prototype evaluations, several routing strategies were evaluated, such as the implementation of routing rules on top of \textit{Twin Instances}, leading to increased RabbitMQ memory usage. In the latest \textit{KTWIN} prototype, the broker instance is optimized to include only two exchanges: one for the broker ingress component, where all Cloud Events are published, and the MQTT exchange, where devices publish and subscribe to messages. The simulation events were effectively routed through only 13 queues regardless of the city size, showcasing \textit{KTWIN}'s scalable and efficient routing capabilities.
}

The number of bindings is proportional to the number of routing rules created by \textit{KTWIN} Operator and defined in the Digital Twin Graph. Bindings are responsible for routing messages from exchanges to queues. The routing mechanism used by \textit{KTWIN} uses the Cloud Events specification header \textit{type}, which includes information about \textit{Twin Interface} that owns that event. In total, for all experiments, 57 bindings were created. The number of bindings created includes different routing rules such as routing events from real devices to virtual \textit{Twin Services}, from \textit{Twin Services} to real devices, from \textit{Twin Services} to related \textit{Twin Services}, or routing events to the event store.

RabbitMQ manages all MQTT connections in a single exchange and uses routing-key bindings to route messages between publishers and subscribers connected to the MQTT exchange, resulting in a binding creation for each subscriber connected to the broker. Because the proposed experiments focused only on the data injection into \textit{KTWIN}, no subscribers were attached to the broker. Simulation with subscriber devices can be part of future search opportunities.

\subsubsection{Services Scalability Analysis}

In this section, we compare the \textit{KTWIN} component's behavior regarding resource usage, auto-scaling and response time for the different Smart City scenarios. The experiments aim to verify whether the prototype implemented can scale for the selected city sizes and identify existing bottlenecks and points of improvement.

\begin{figure}[ht]
    \centering
    \includegraphics[scale=0.45]{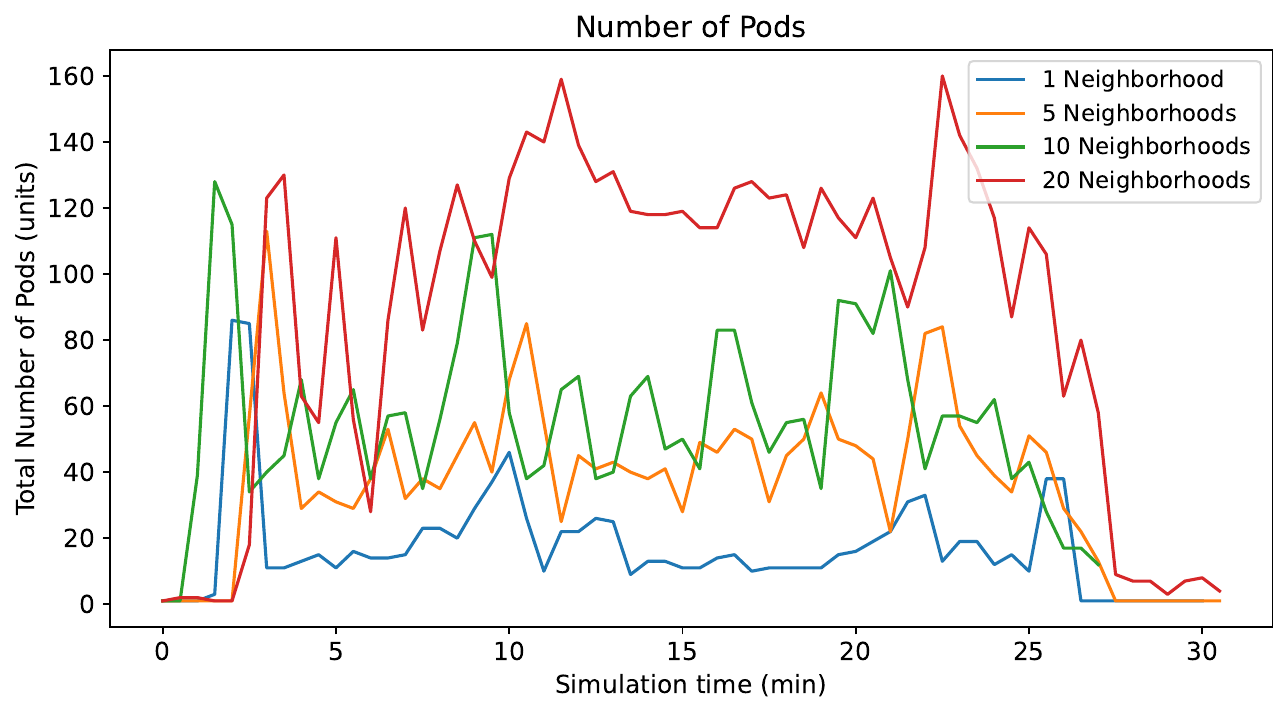}
    \caption{The number of pods instantiated by auto scaler during the simulations.}
    \label{fig:evaluation:experiment-2:total-number-of-pods}
\end{figure}

Figure \ref{fig:evaluation:experiment-2:total-number-of-pods} shows the total number of pods created during the simulation timeframe. In all experiments' first burst of requests, the number of pods is close to zero because scale-to-zero was enabled. As soon as the first requests are triggered and buffered in \textit{Knative} \textit{Activator}, \textit{KTWIN} scales up the number of replicas so messages can be processed as soon as possible. The chart shows that the total number of pods instantiated increases as the experiment size increases.  With the configured auto-scaling settings, the mean number of pods created for the experiments of 1, 5, 10 and 20 neighborhoods was 17, 37, 55 and 88 respectively, reaching a maximum of 86, 113, 128 and 160 pods per each experiment. The auto-scaling metric adopted for all experiments was the Knative concurrency metric with a value of 5. The concurrency metric determines the number of simultaneous requests that each replica of an application can process at any given time. 

\begin{figure}[h]
\begin{subfigure}{0.49\textwidth}
    \includegraphics[scale=0.38]{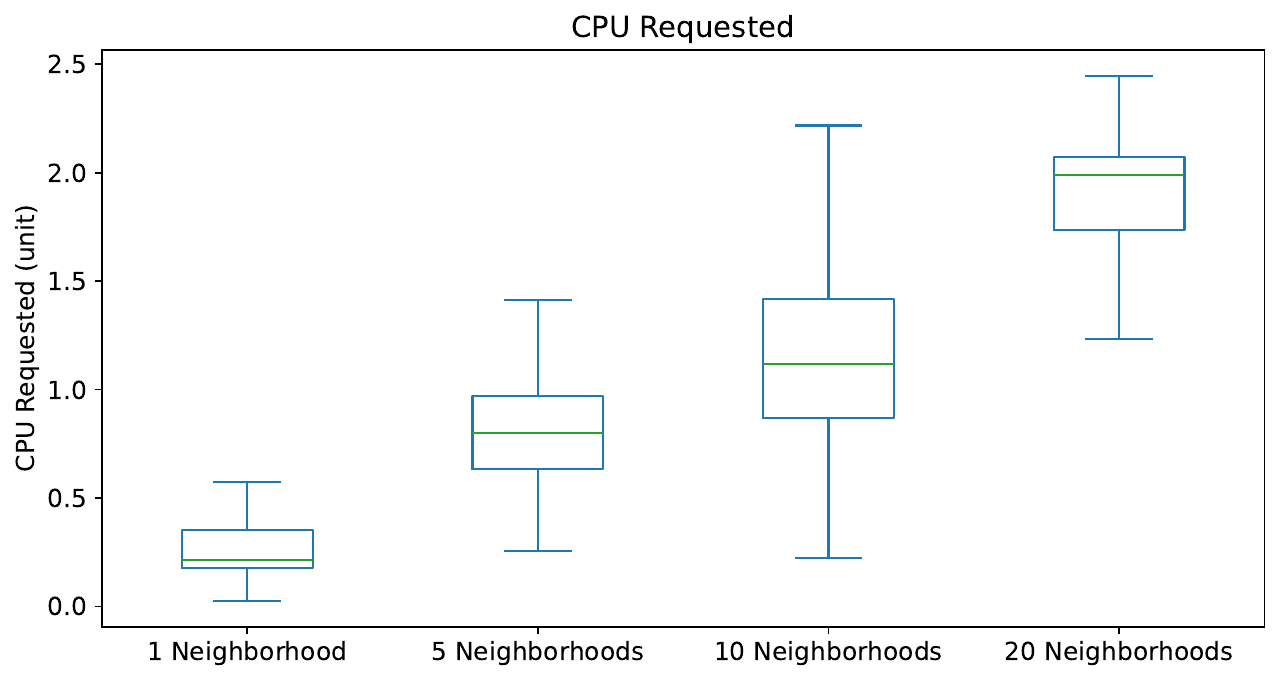}
    \caption{The average distribution of CPU units requested by services.}
    \label{fig:evaluation:experiment-2:cpu-requested-services}
\end{subfigure}
\begin{subfigure}{0.49\textwidth}
    \includegraphics[scale=0.38]{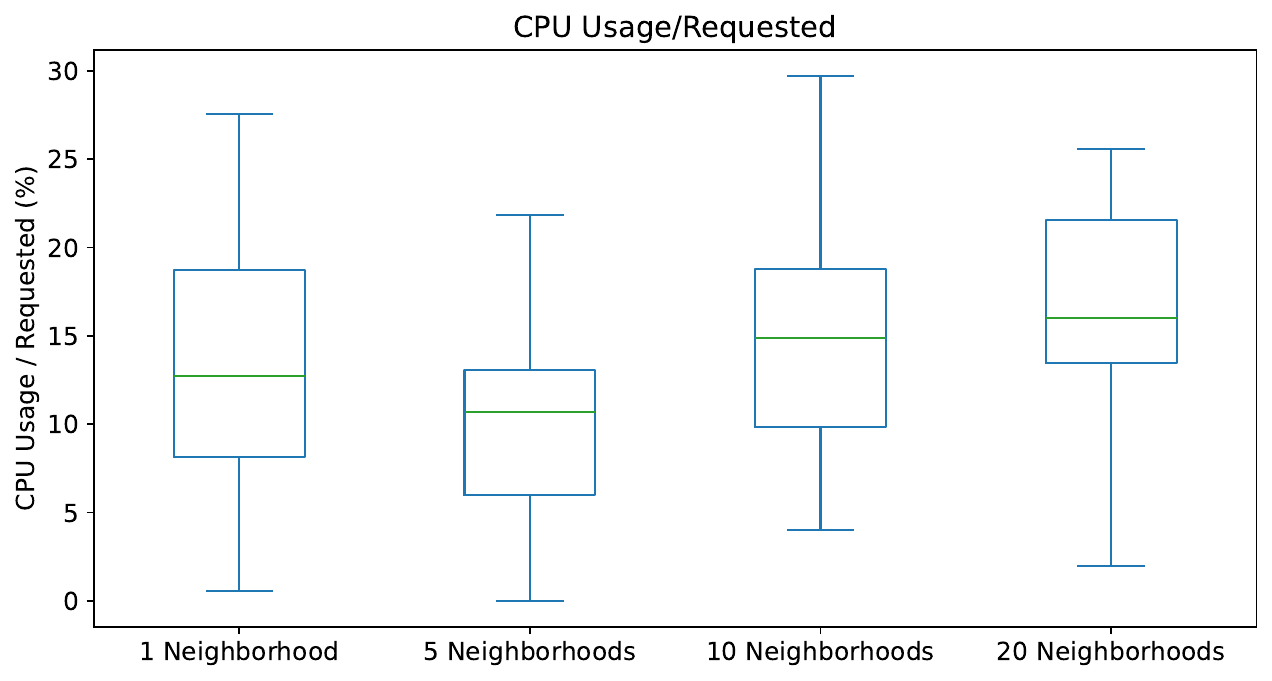}
    \caption{The average distribution of CPU percentage used by services.}
    \label{fig:evaluation:experiment-2:cpu-usage-services}
\end{subfigure}

\caption{The average of CPU requested and used by services.}
\label{fig:evaluation:experiment-2:cpu-usage-requested-services}
\end{figure}

The number of CPU resources allocated by \textit{KTWIN} services also increases as the number of events to be processed increases. Figure \ref{fig:evaluation:experiment-2:cpu-usage-requested-services} depicts the average distribution of the CPU requested and the percentage of CPU used by all \textit{Twin Services}. In Figure \ref{fig:evaluation:experiment-2:cpu-requested-services}, the requested CPU value is higher in larger scenarios because it is proportional to the number of pods created to process requests: each pod created requests 0.1 CPU for Kubernetes. In Figure \ref{fig:evaluation:experiment-2:cpu-usage-services}, the results show that for the conducted experiments, the CPU percentage usage was above 10\% and below 30\% during most of the time of the experiments, which demonstrates that the system was able to respond to variations in the load of events for all scenarios, consuming a similar range of computing resource even with a different number of instances.

\begin{figure}[ht]
\begin{subfigure}{0.49\textwidth}
    \includegraphics[scale=0.37]{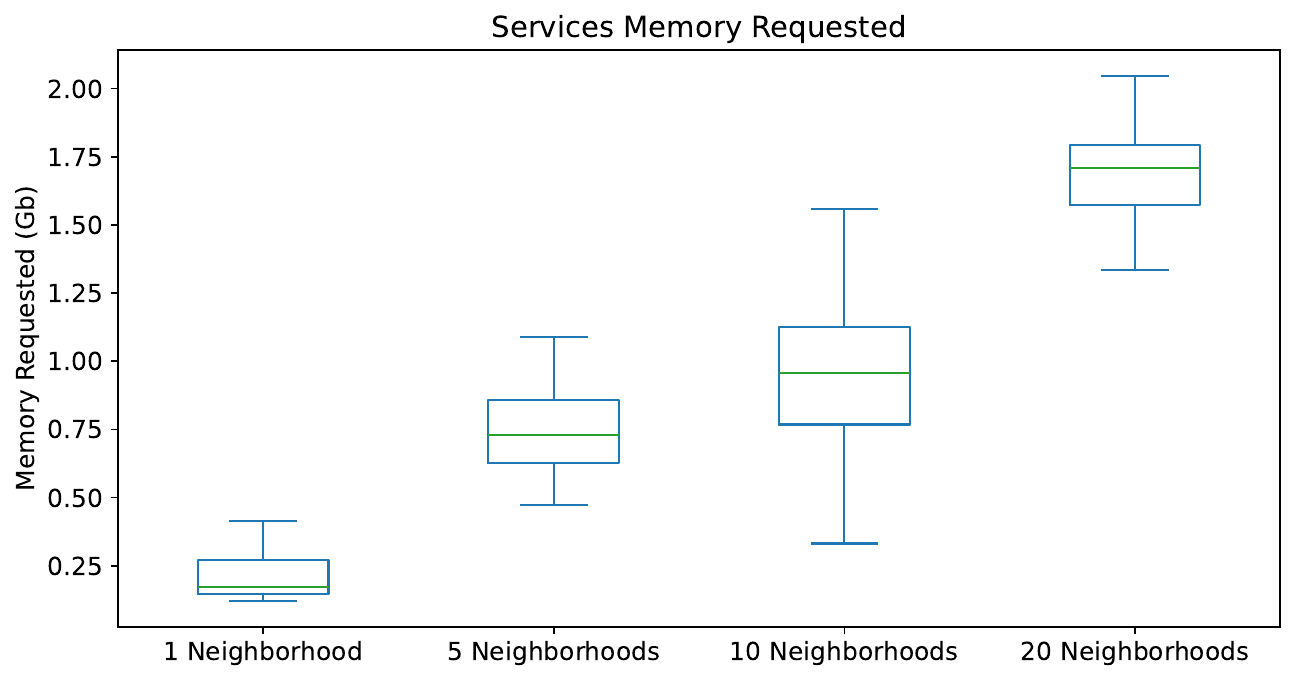}
    \caption{The average of memory requested by services.}
    \label{fig:evaluation:experiment-2:memory-requested-services}
\end{subfigure}
\begin{subfigure}{0.49\textwidth}
    \includegraphics[scale=0.37]{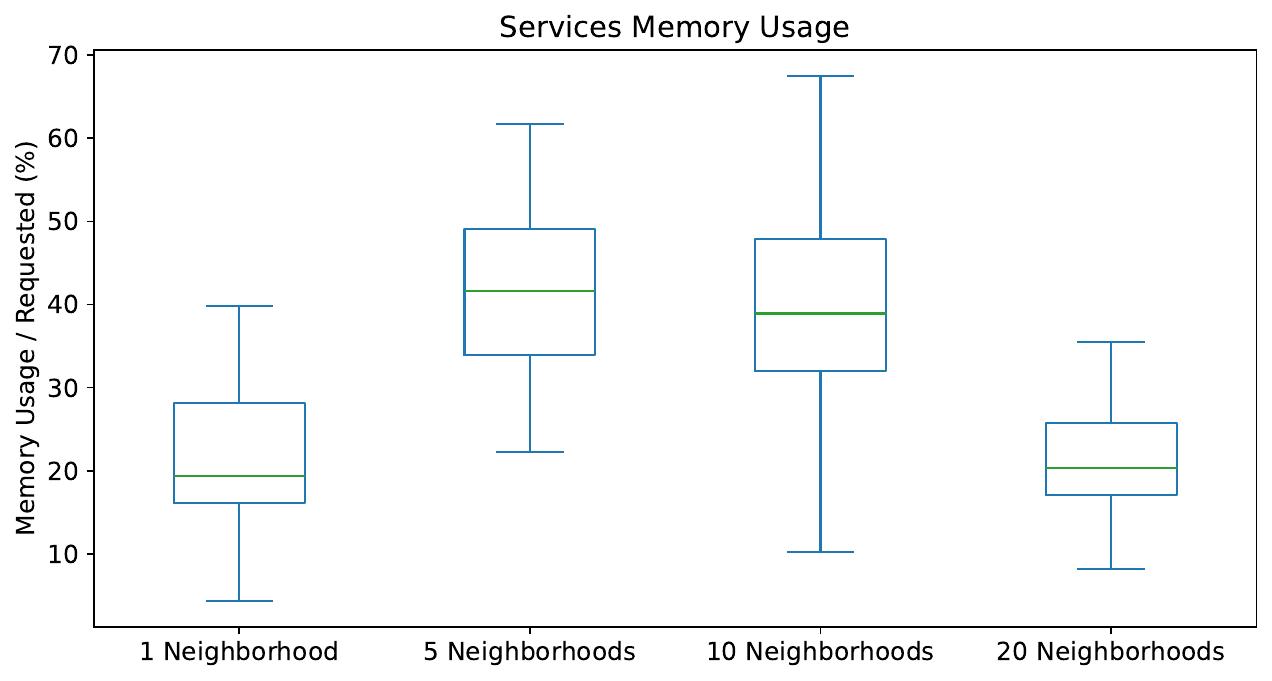}
    \caption{The average of memory used by services.}
    \label{fig:evaluation:experiment-2:memory-usage-services}
\end{subfigure}

\caption{The average of memory requested and used by services.}
\label{fig:evaluation:experiment-2:memory-usage-requested-services}
\end{figure}

Memory usage is also part of this analysis. Figure \ref{fig:evaluation:experiment-2:memory-requested-services} shows the distribution of requested memory in Gigabytes: as the amount of pods grows, the amount of allocated memory increases. In Figure \ref{fig:evaluation:experiment-2:memory-usage-services}, it can be verified that for different experiment sizes, memory usage remains between the range of 15\% and 50\% during at least half of the execution time, showing that the system can respond well to different event bursts. This chart also shows that the distribution range of memory usage reduces for larger experiments, indicating a higher amount of allocated resources than required.

\begin{figure}[!]
\begin{subfigure}{0.5\textwidth}
    \includegraphics[scale=0.38]{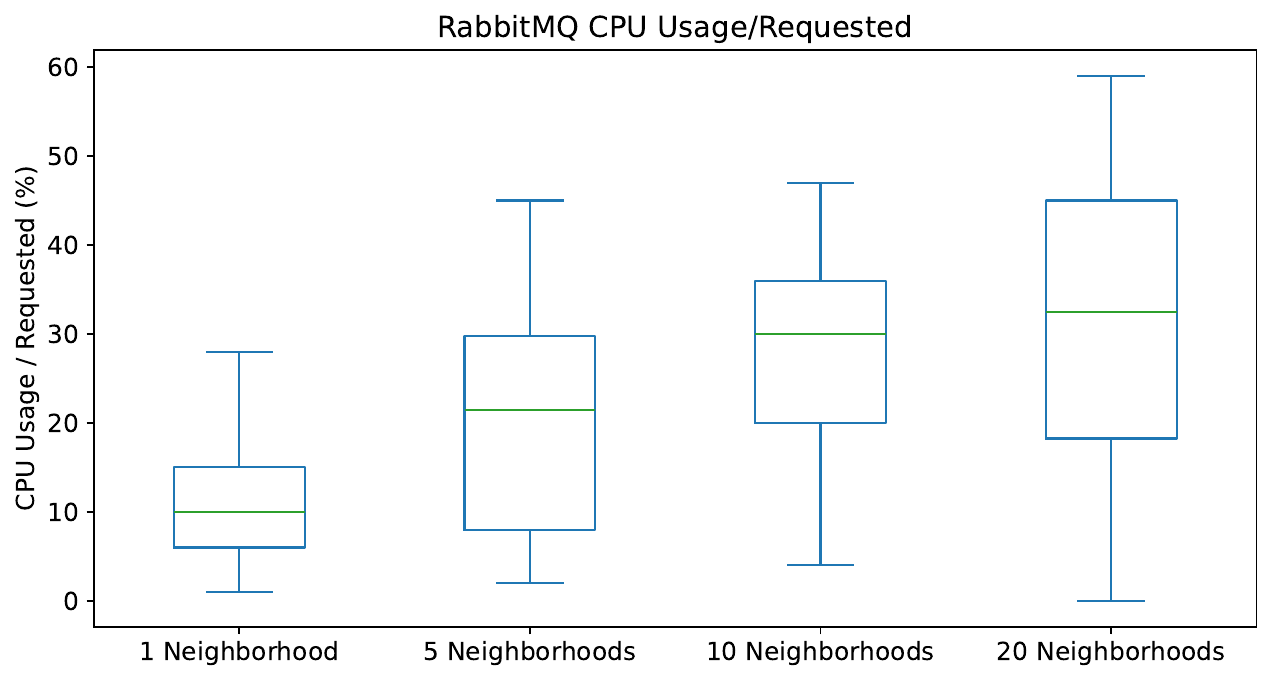}
    \caption{RabbitMQ used/requested by CPU.}
    \label{fig:evaluation:experiment-2:cpu-usage-core-rabbitmq}
\end{subfigure}
\begin{subfigure}{0.5\textwidth}
    \includegraphics[scale=0.38]{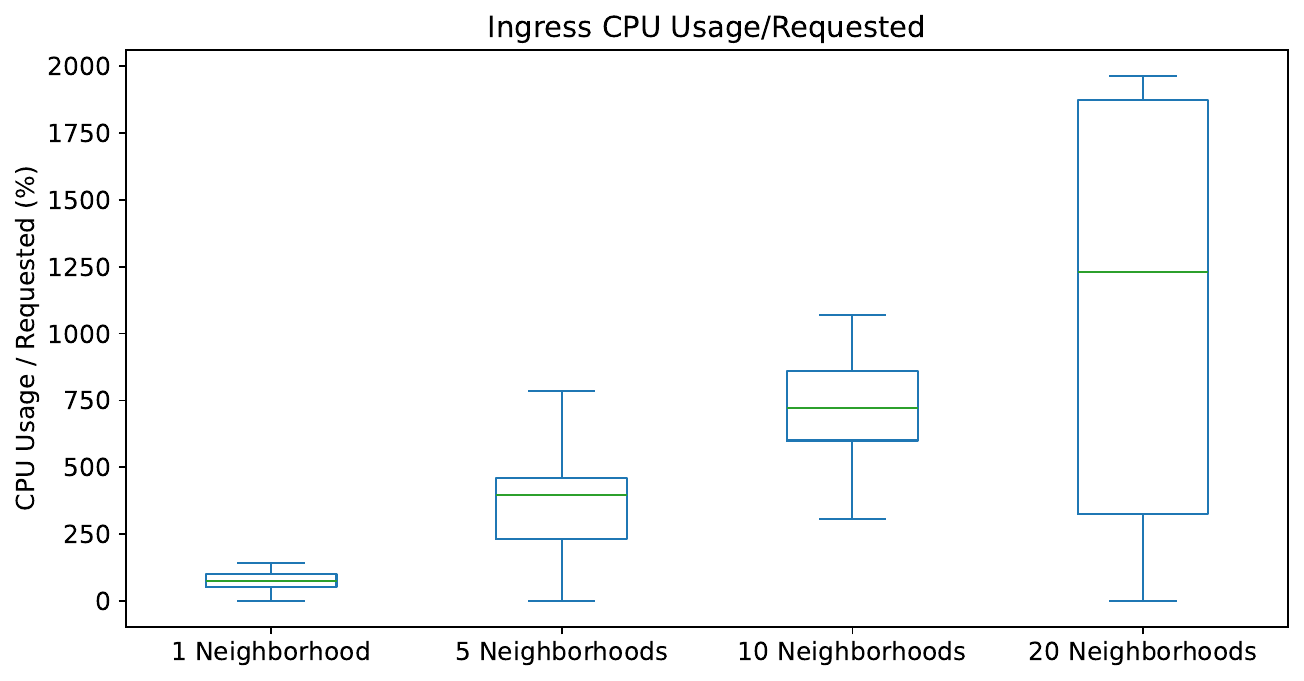}
    \caption{Broker Ingress used/requested CPU.}
    \label{fig:evaluation:experiment-2:cpu-usage-core-ingress}
\end{subfigure}
\begin{subfigure}{0.5\textwidth}
    \includegraphics[scale=0.38]{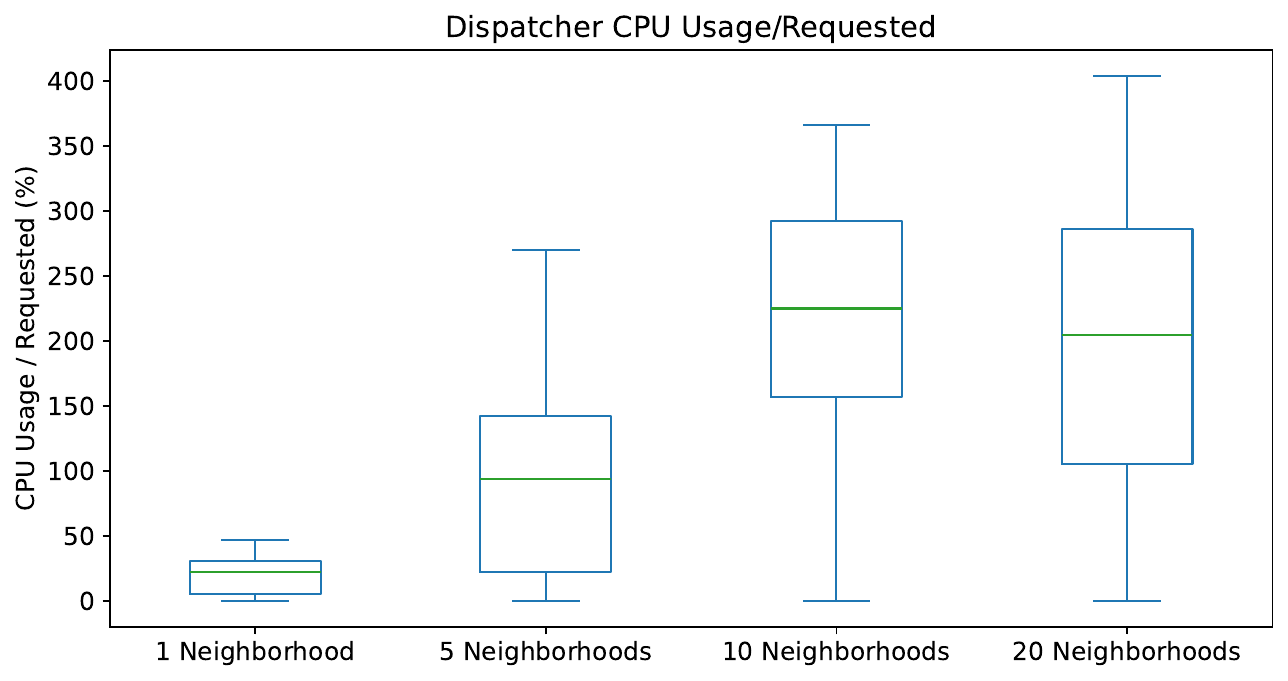}
    \caption{Dispatcher's CPU used/requested.}
    \label{fig:evaluation:experiment-2:cpu-usage-core-dispatcher}
\end{subfigure}
\begin{subfigure}{0.5\textwidth}
    \includegraphics[scale=0.38]{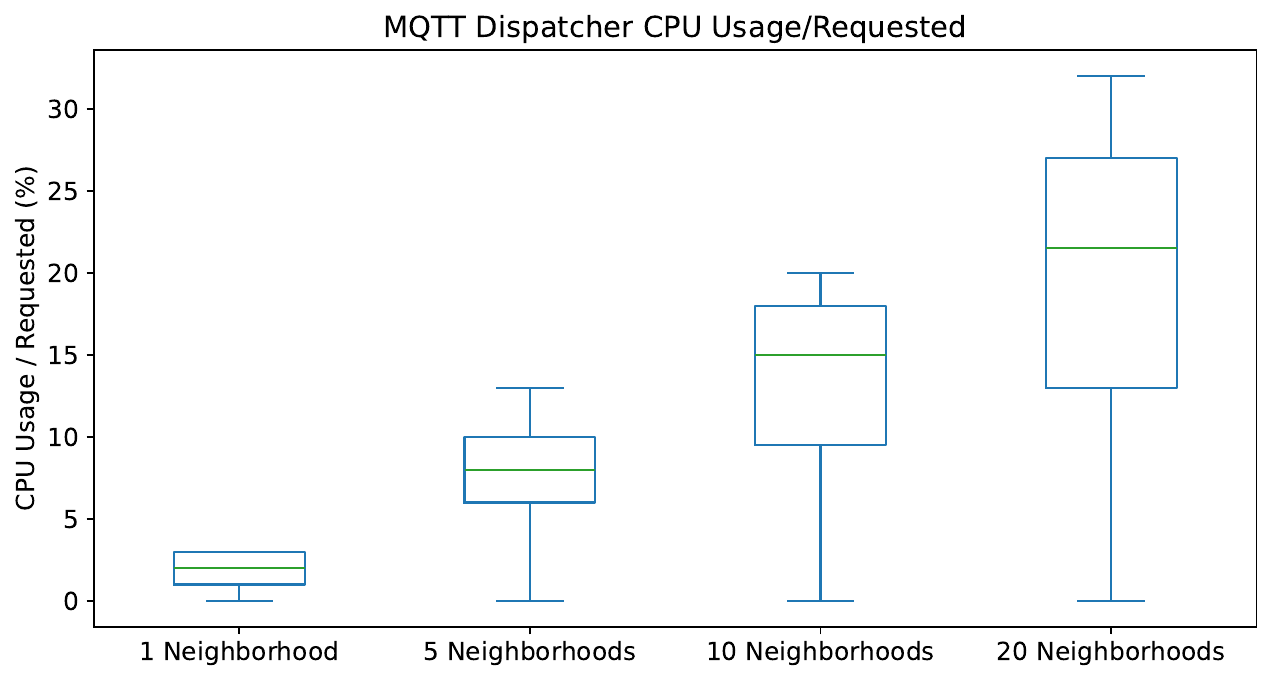}
    \caption{MQTT Dispatcher used/requested CPU.}
    \label{fig:evaluation:experiment-2:cpu-usage-core-mqtt-dispatcher}
\end{subfigure}

\caption{The distribution of used CPU by core components.}
\label{fig:evaluation:experiment-2:cpu-usage-requested-core}
\end{figure}

Also regarding CPU usage, Figure \ref{fig:evaluation:experiment-2:cpu-usage-requested-core} shows the distribution of CPU utilization for some of the core components of \textit{KTWIN}. Because some of the core components do not have any auto-scaling policy implemented in these experiments, the CPU usage increases as the experiment size increases. This behavior can be verified for RabbitMQ, Ingress, Dispatchers and MQTT Dispatcher. In the Ingress and Dispatcher charts, the CPU usage exceeds 100\% of the requested amount, especially for the largest city sizes. This indicates that the corresponding applications were running with under-provisioned resources, and it also represents a point of improvement for future research, such as implementing auto-scaling features for those core components. 


\begin{figure}[!]
\begin{subfigure}{0.5\textwidth}
    \includegraphics[scale=0.38]{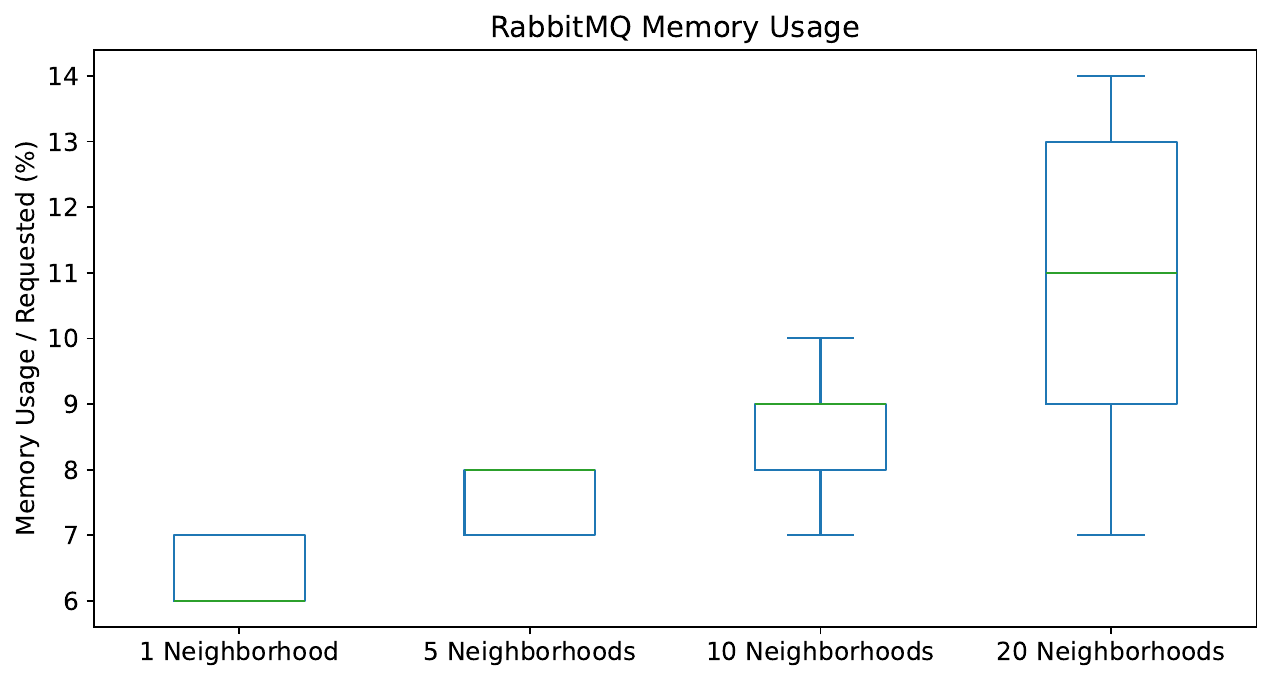}
    \caption{RabbitMQ used/requested memory .}
    \label{fig:evaluation:experiment-2:memory-usage-core-rabbitmq}
\end{subfigure}
\begin{subfigure}{0.5\textwidth}
    \includegraphics[scale=0.38]{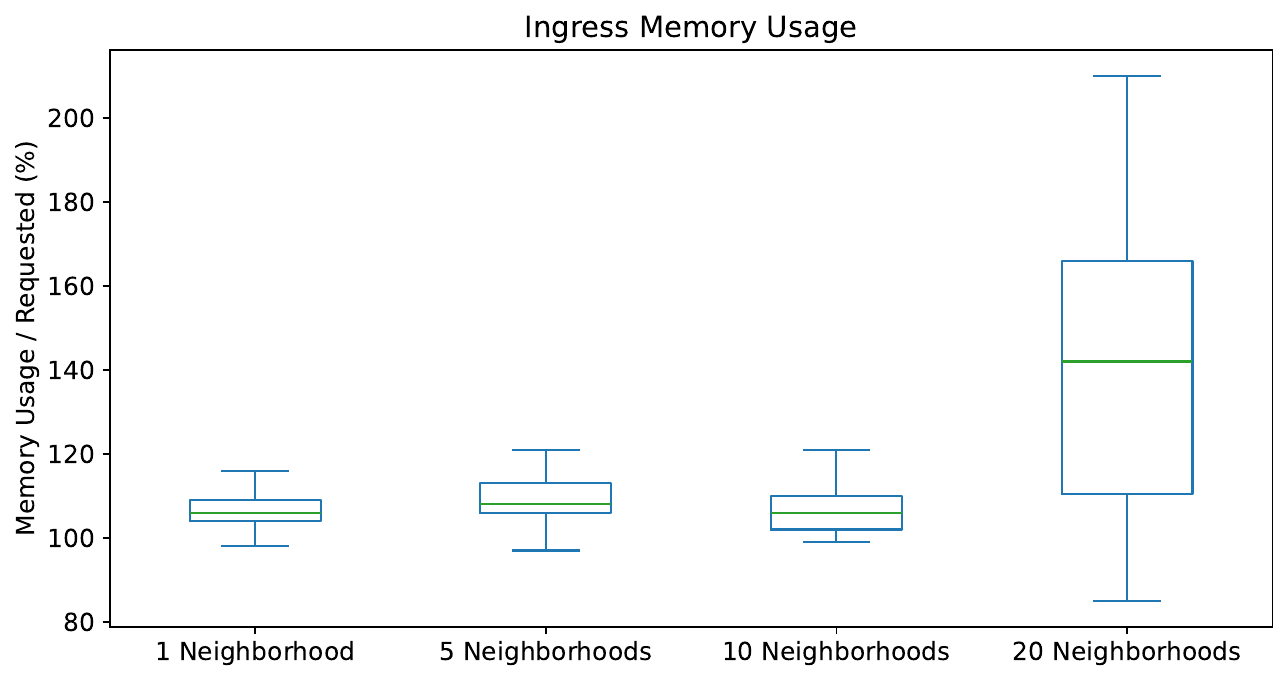}
    \caption{Ingress used/requested memory.}
    \label{fig:evaluation:experiment-2:memory-usage-core-ingress}
\end{subfigure}
\begin{subfigure}{0.5\textwidth}
    \includegraphics[scale=0.38]{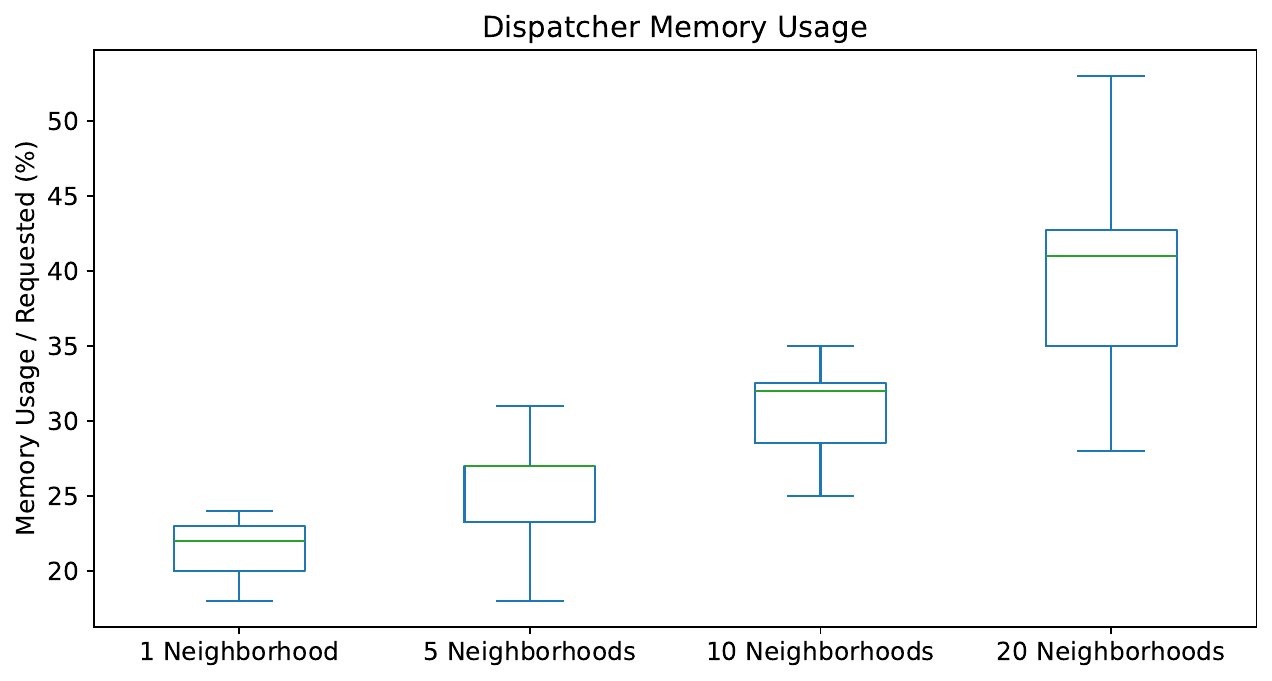}
    \caption{Dispatcher's used/requested memory .}
    \label{fig:evaluation:experiment-2:memory-usage-core-dispatcher}
\end{subfigure}
\begin{subfigure}{0.5\textwidth}
    \includegraphics[scale=0.38]{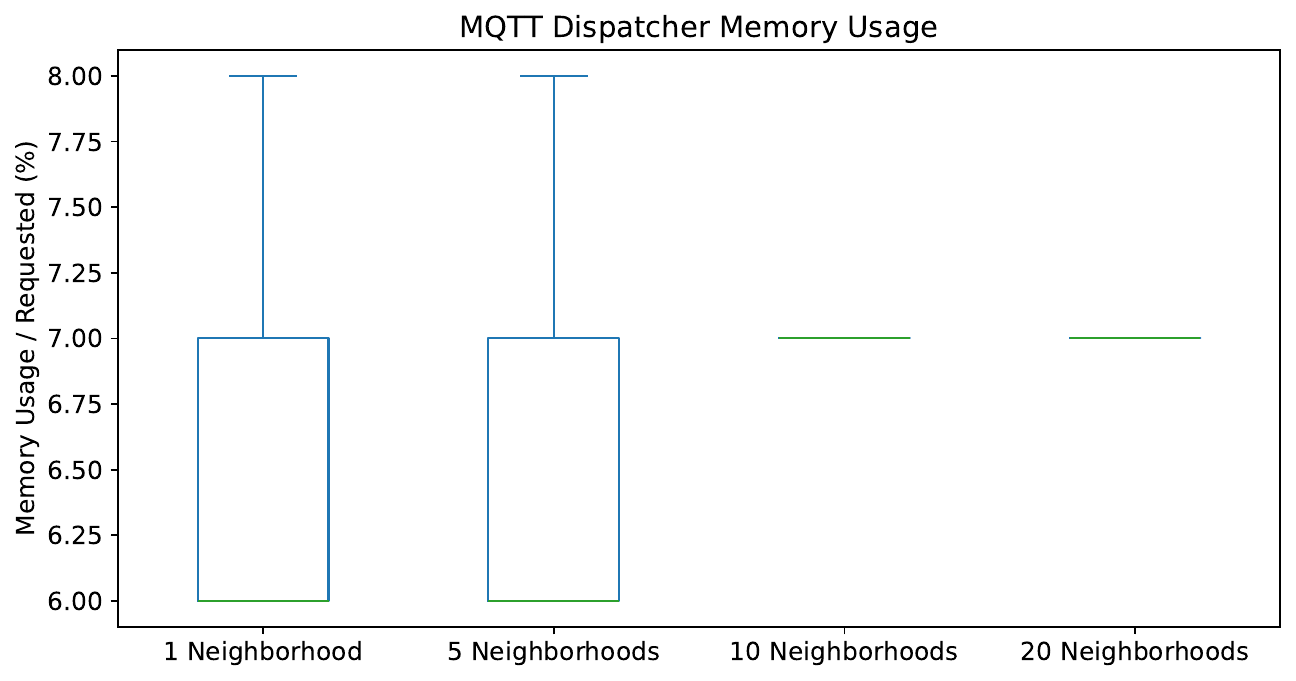}
    \caption{MQTT Dispatcher used/requested memory.}
    \label{fig:evaluation:experiment-2:memory-usage-core-mqtt-dispatcher}
\end{subfigure}

\caption{The distribution of used memory by core components.}
\label{fig:evaluation:experiment-2:memory-usage-requested-core}
\end{figure}

Figure \ref{fig:evaluation:experiment-2:memory-usage-requested-core} shows that the memory percentage usage of \textit{KTWIN} core components also increases as the experiment size for most of the experiments. The memory remains constant for the MQTT Dispatcher component. In contrast to the CPU usage percentage, the amount of memory required by Redis increases for larger experiments mostly because it requires more space to cache the Twin Graph size. Finally, the Event Store Service memory consumption indicates that as larger the experiment was, \textit{KTWIN} instantiated the required number of pods to handle the published events.

\begin{figure}[ht]
\begin{subfigure}{0.5\textwidth}
    \centering
    \includegraphics[scale=0.45]{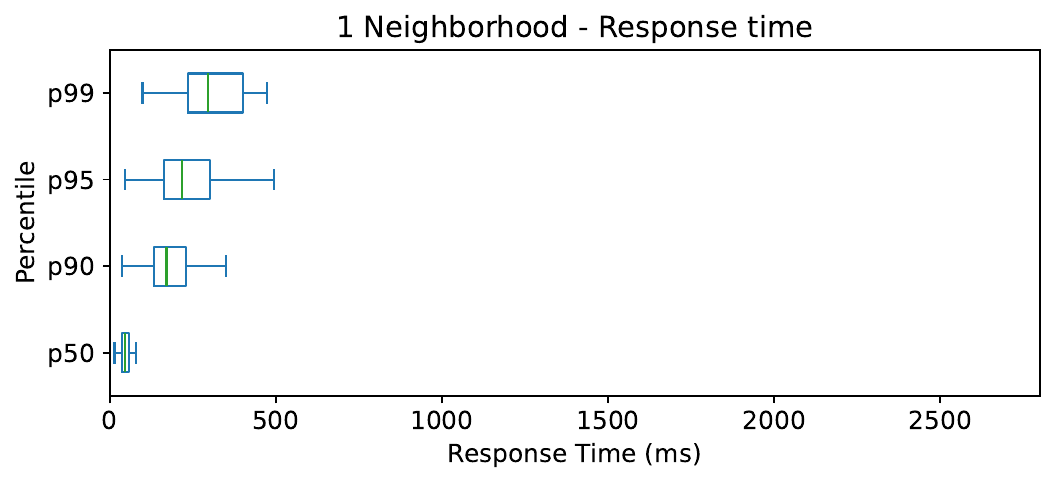}
    \caption{1 Neighborhood-scenario}
    \label{fig:evaluation:experiment-2:response-time-1-neighborhood}
\end{subfigure}
\begin{subfigure}{0.5\textwidth}
    \centering
    \includegraphics[scale=0.45]{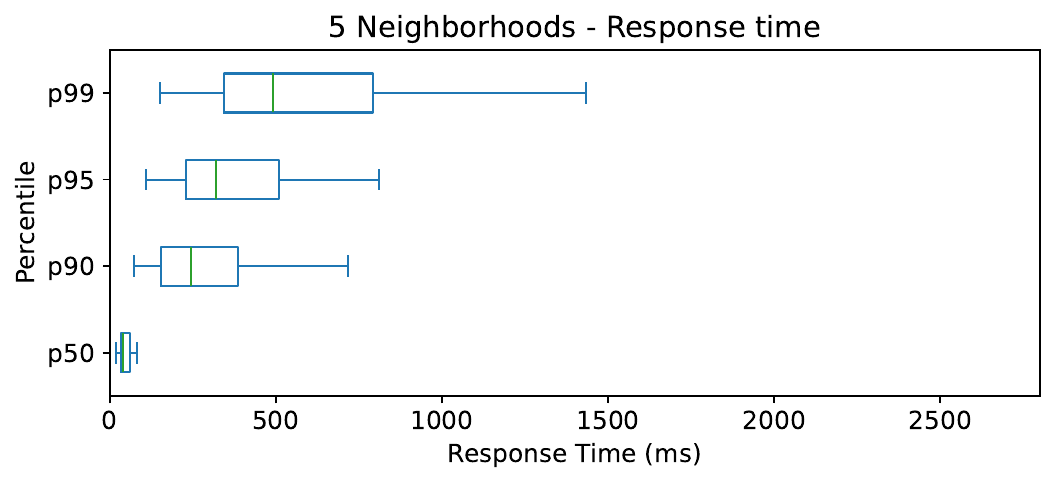}
    \caption{5 Neighborhoods-scenario}
    \label{fig:evaluation:experiment-2:response-time-5-neighborhood}
\end{subfigure}
\begin{subfigure}{0.5\textwidth}
    \centering
    \includegraphics[scale=0.45]{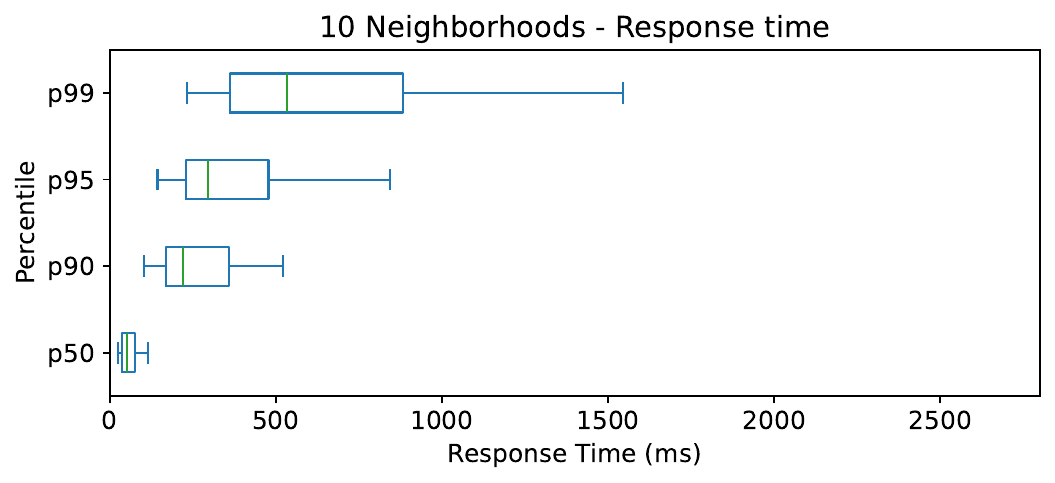}
    \caption{10 Neighborhoods-scenario}
    \label{fig:evaluation:experiment-2:response-time-10-neighborhood}
\end{subfigure}
\begin{subfigure}{0.5\textwidth}
    \centering
    \includegraphics[scale=0.45]{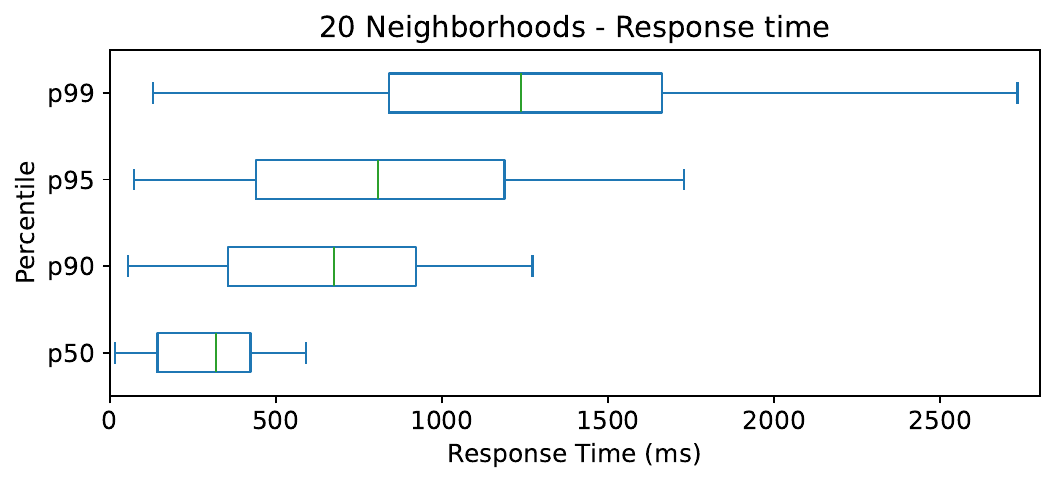}
    \caption{20 Neighborhoods-scenario}
    \label{fig:evaluation:experiment-2:response-time-20-neighborhood}
\end{subfigure}
    \caption{The Twin Services average response time during the experiments.}
    \label{fig:evaluation:experiment-2:response-time-services}
\end{figure}

Finally, Figure \ref{fig:evaluation:experiment-2:response-time-services} shows the distribution average response time of \textit{Twin Services} for each scenario. The chart divides the responses into percentile 50th (p50), percentile 90th (p90), percentile 95th (p95), and percentile 99th (p99). We expect that the larger the city size is, the more requests are processed, hence the greater the response time. This fact can be verified if we compare the 1-neighborhood scenario against the other configurations. The similarities in the response time between the 5-neighborhood and 10-neighborhood scenarios indicate that the system resolves requests in the same period, while the amount of requests is multiplied by two. When analyzing the larger city scenario, we verify the application response time increases by some orders of magnitude. The boxplot also shows some outlier points in all experiment scenarios, indicating the impact of the cold start on the response time.

The conducted experiments conclude that \textit{KTWIN} system components scale very well for the selected use case. With the hardware available for the experiments the \textit{Event Broker} achieves up to 1800 requests per second without presenting resource usage issues. The system bottleneck was achieved at the threshold of 350 requests per second, where the number of events per second was higher than the number of records that could be processed by Knative dispatchers, resulting in some messages being queued for some time before being dispatched to the \textit{Twin Services}. The identified bottleneck can be explained because dispatchers do not contain auto-scaling and work with a limited and static amount of requested CPU and memory. This limitation may be part of future work improvement opportunities. Nevertheless, because the designed experiments represent a hypothetical day in a modern city in a short period of 24 minutes, a 60 times shorter period compared to a full 24-hour day, we can conclude that the implemented prototype should present similar results for cities 60 times larger in case the event intervals also increase in the same rate.

\subsubsection{Serverless Capabilities Analysis}


\textit{KTWIN} Serverless capabilities provide several benefits to end users. The platform provides operational complexity and cost reduction allowing domain experts to define their own ontology-based Digital Twin definitions and perform rapid development of \textit{Twin Interface} functions without requiring deep knowledge of infrastructure management. \textit{KTWIN} Serverless capabilities also include the possibility to offer the solution as a Pay-as-you-Go subscription model in which users only pay for the compute time, the number of events processed, the twin size and the amount of data storage. In addition, \textit{KTWIN} provides parameterized scalability of components with automatic scaling based on demand and the dynamic allocation of resources to ensure optimal performance during varying loads. These aspects enable \textit{KTWIN} to be the foundation of a Serverless Platform for Digital Twins. In order to evaluate the functions' scalability and the achievable cost savings, we compare the 1-neighborhood and 5-neighborhoods experiments against scenarios with fixed amounts of under-provisioned resources (the number of pods per \textit{Twin Interface} was set to 1), and over-provisioned scenarios (14 pods per \textit{Twin Interface}).

\begin{figure}[ht]
    \centering
    \includegraphics[scale=0.40]{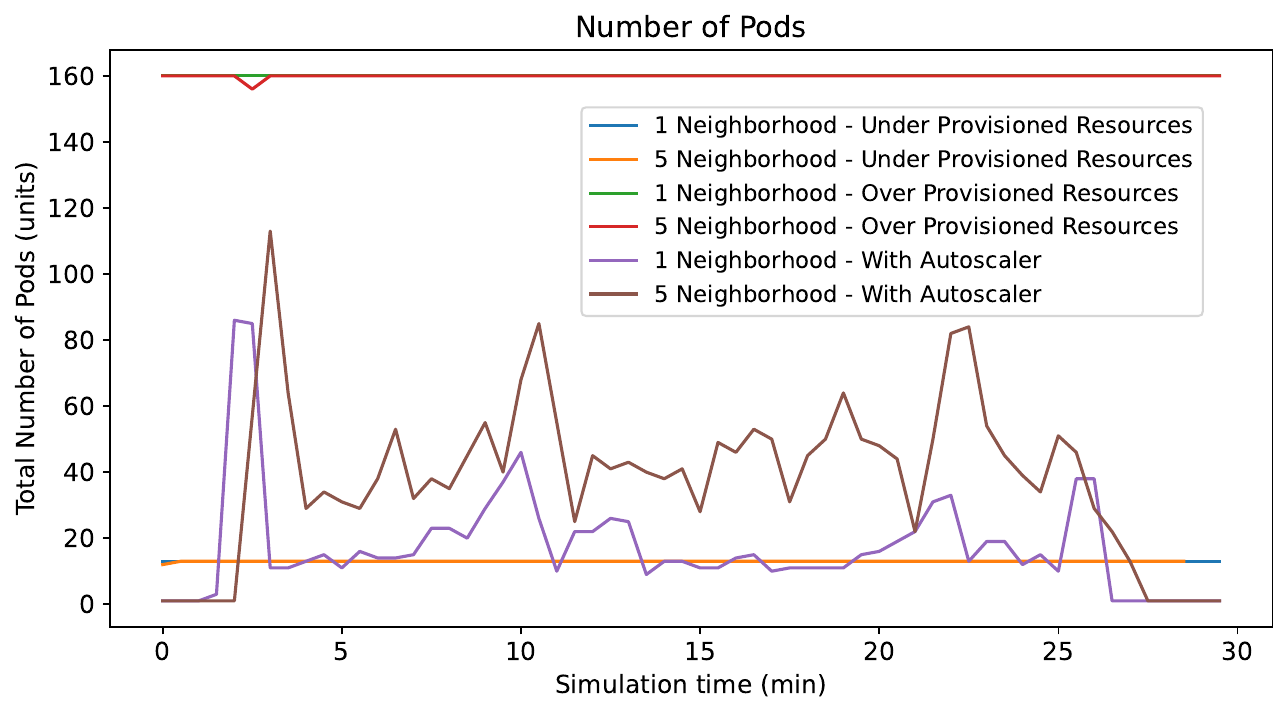}
    \caption{The number of pods instantiated by auto scaler during the simulations.}
    \label{fig:evaluation:experiment-3:total-number-of-pods}
\end{figure}

Figure \ref{fig:evaluation:experiment-3:total-number-of-pods} depicts the number of pods instantiated during the simulation period. The total number of pods of the over-provisioned scenarios has a constant line of 160, while the under-provisioned has a fixed line of 13 pods. The intermediary lines represent the scenarios with auto-scaling enabled. The mean number of instances in the 1-neighborhood scenario is 17, while in the 5-neighborhood scenario, the average is 38.

\begin{figure}[ht]
\begin{subfigure}{0.5\textwidth}
    \centering
    \includegraphics[scale=0.40]{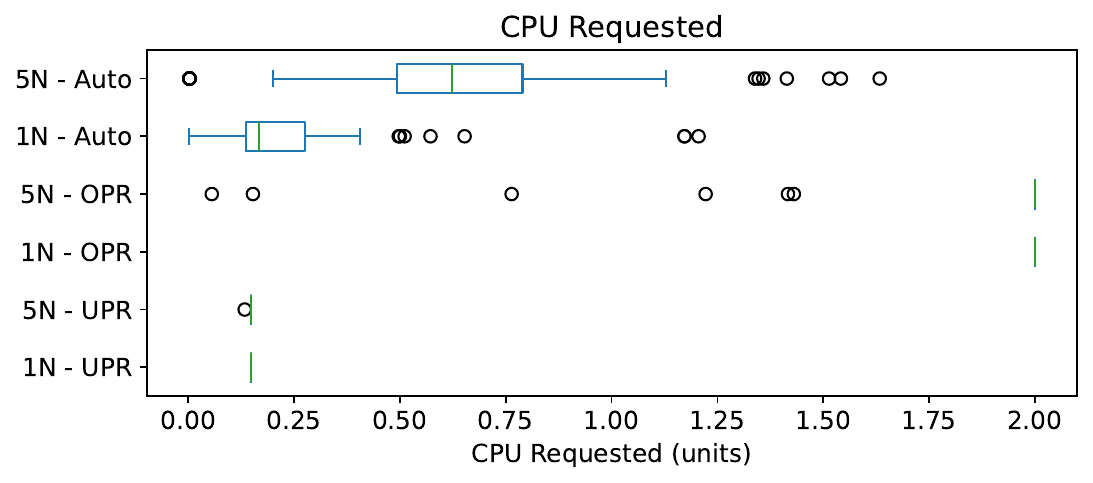}
    \caption{Services CPU requested.}
    \label{fig:evaluation:experiment-3:cpu-requested-services}
\end{subfigure}
\begin{subfigure}{0.5\textwidth}
    \centering
    \includegraphics[scale=0.40]{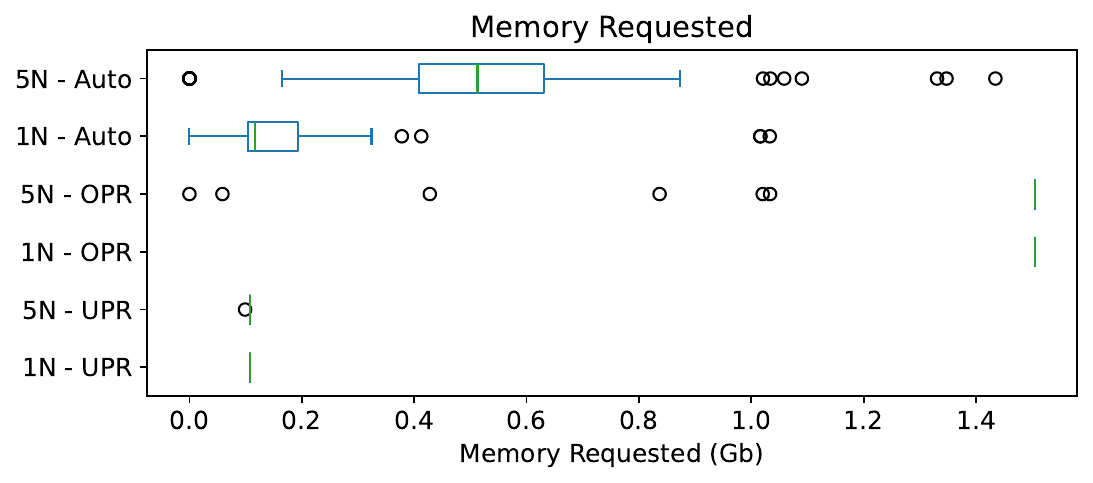}
    \caption{Services memory requested.}
    \label{fig:evaluation:experiment-3:memory-requested-services}
\end{subfigure}
\begin{subfigure}{0.5\textwidth}
    \centering
    \includegraphics[scale=0.40]{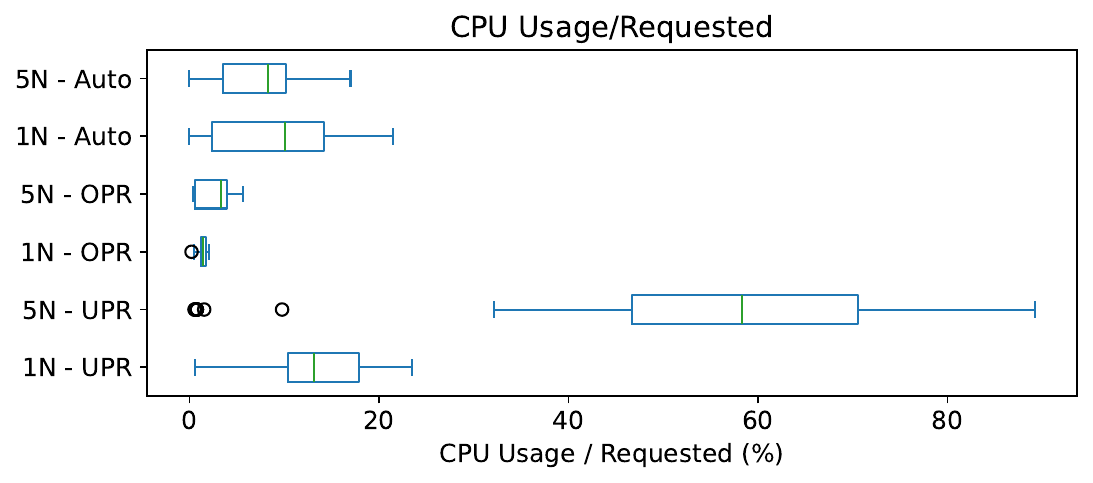}
    \caption{Services CPU used/requested.}
    \label{fig:evaluation:experiment-3:cpu-usage-services}
\end{subfigure}
\begin{subfigure}{0.5\textwidth}
    \centering
    \includegraphics[scale=0.40]{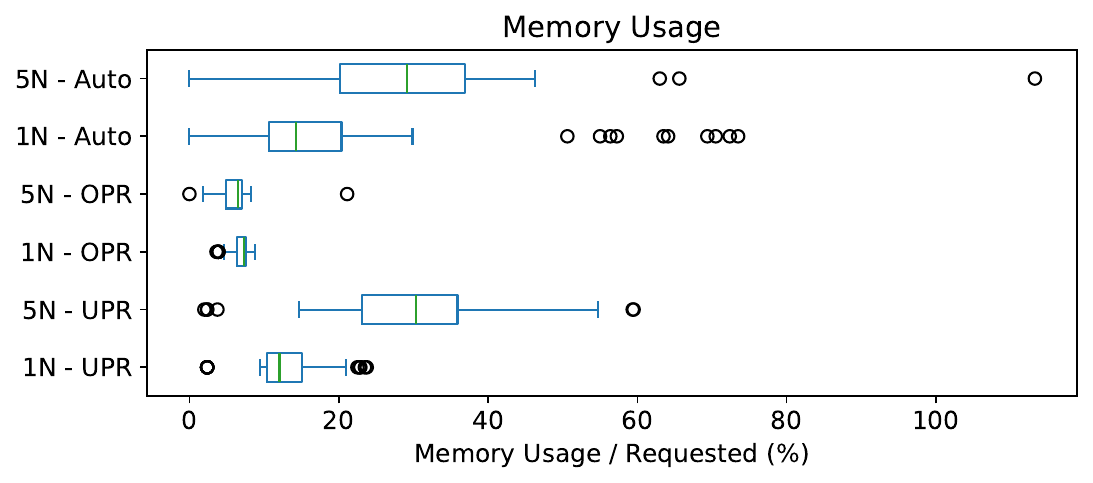}
    \caption{Services memory used/requested.}
    \label{fig:evaluation:experiment-3:memory-usage-services}
\end{subfigure}
\caption{CPU requested and used by services.}
\label{fig:evaluation:experiment-3:cpu-usage-requested-services}
\end{figure}

The savings in resource usage become evident when we analyze the allocated and used CPU and memory in the simulated scenarios. In Figure \ref{fig:evaluation:experiment-3:cpu-requested-services}, we can verify that for the scenarios with a fixed number of pods, the amount of allocated CPU remains the same, while in the scenarios with automatic scale enabled, the amount of requested CPU changes due to the variance in the number of instances created. Although some of the \textit{Twin Instances} are scaled to the maximum of 15 pods in some moments of the experiment, the requested CPU of automatic scale scenarios remains beneath the scenarios with over-provisioned, representing potential savings between 67\% and 88\% in allocated CPU costs.

Figure \ref{fig:evaluation:experiment-3:cpu-usage-services} shows that the CPU used by the services for the scenarios with automatic scale remains under 20\% of usage, almost the same range registered in the overprovisioned scenario. In contrast, CPU usage in the under-provisioned 5-neighborhood scenario stays above 50\% during most of the experiment period. The chart presented in \ref{fig:evaluation:experiment-3:cpu-usage-services} suggests that auto-scaler scenarios can allocate less resource usage and use the allocated ones more efficiently.


The same resource usage efficiency can be verified in the memory usage charts. Figures \ref{fig:evaluation:experiment-3:memory-requested-services} and \ref{fig:evaluation:experiment-3:memory-usage-services} show that the amount of allocated memory remains far below the scenarios of over-provisioned resources, while the percentage of usage memory presents a better efficiency compared to the under-provisioned scenario. The allocated memory savings range between 64\% and 89\% compared to the over-provisioned configuration.

\begin{figure}[h!]
\begin{subfigure}{0.5\textwidth}
    \centering
    \includegraphics[scale=0.38]{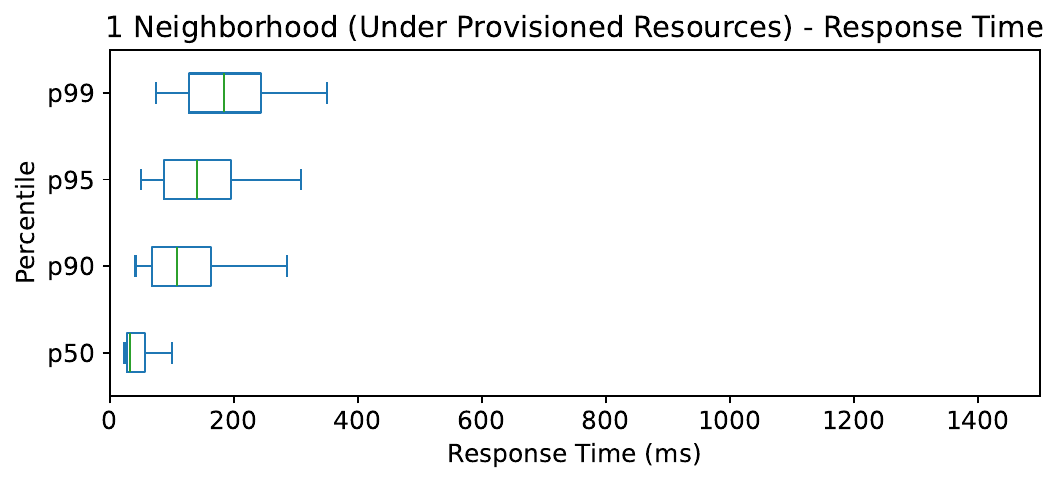}
    \caption{1 Neighborhood under-provisioned scenario}
    \label{fig:evaluation:experiment-3:response-time-1-neighborhood-under-provisioned}
\end{subfigure}
\begin{subfigure}{0.5\textwidth}
    \centering
    \includegraphics[scale=0.38]{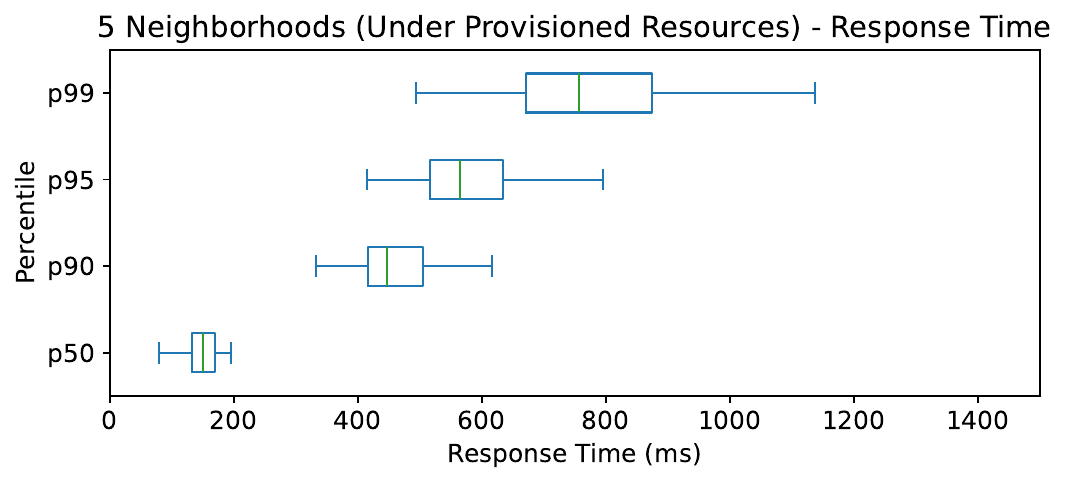}
    \caption{5 Neighborhood under-provisioned scenario}
    \label{fig:evaluation:experiment-3:response-time-5-neighborhood-under-provisioned}
\end{subfigure}
\begin{subfigure}{0.5\textwidth}
    \centering
    \includegraphics[scale=0.38]{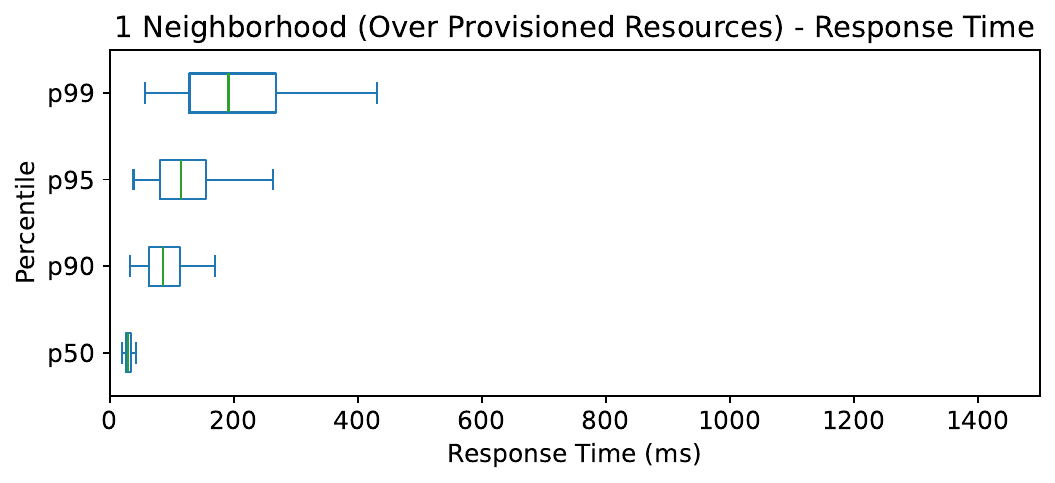}
    \caption{1 Neighborhood over-provisioned scenario}
    \label{fig:evaluation:experiment-3:response-time-1-neighborhood-over-provisioned}
\end{subfigure}
\begin{subfigure}{0.5\textwidth}
    \centering
    \includegraphics[scale=0.38]{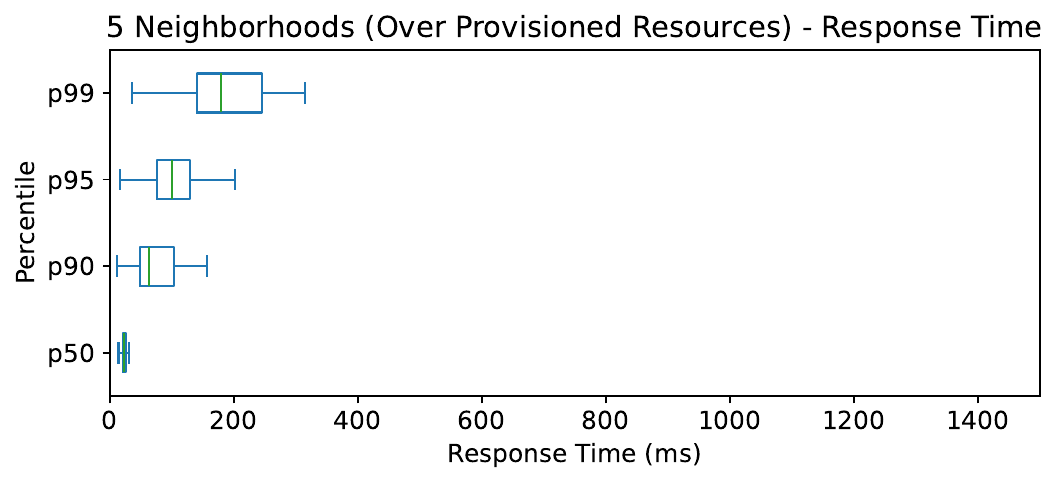}
    \caption{5 Neighborhood over-provisioned scenario}
    \label{fig:evaluation:experiment-3:response-time-5-neighborhood-over-provisioned}
\end{subfigure}
\begin{subfigure}{0.5\textwidth}
    \centering
    \includegraphics[scale=0.38]{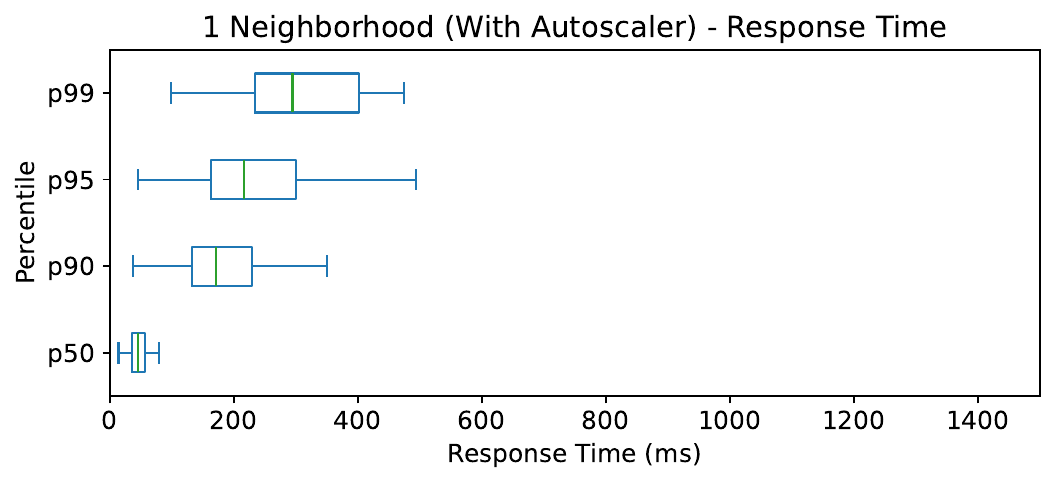}
    \caption{1 Neighborhood with auto-scaling scenario}
    \label{fig:evaluation:experiment-3:response-time-1-neighborhood-with-autoscaler}
\end{subfigure}
\begin{subfigure}{0.5\textwidth}
    \centering
    \includegraphics[scale=0.38]{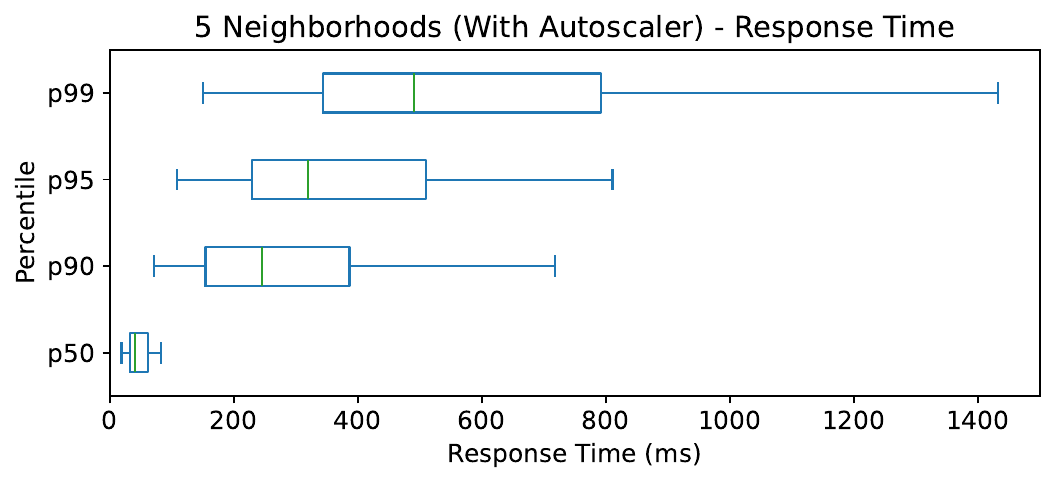}
    \caption{5 Neighborhood with auto-scaling scenario}
    \label{fig:evaluation:experiment-3:response-time-5-neighborhood-with-autoscaler}
\end{subfigure}
    \caption{The average response time of services.}
    \label{fig:evaluation:experiment-3:response-time-services}
\end{figure}

The better efficiency in CPU and memory usage contributes to better system performance, resulting in a reduced response time. Figure \ref{fig:evaluation:experiment-3:response-time-services} shows that the responses in the scenarios with automatic scale, represented by the 50th percentile (p50) boxplot, present a faster response time than the over-provisioned scenario when comparing cities of the same size. Still, when comparing the under-provisioned and auto-scaling scenarios, the median response time has fewer impacts as the system scales in size, indicating that the auto-scaling setup can respond better to higher bursts of data. In contrast, the 99th percentile chart also suggests the auto-scaling scenario response time has impacts due to the cold start issue, especially because the p99 box is more spread compared to the other configurations. As expected, the scenarios with over-provisioned resources present the best response times compared to the other configurations.

The results show that \textit{KTWIN} achieves resource usage efficiency closer to the under-provisioned scenario, while still providing performance near the over-provisioned scenario. In contrast, the scenarios with autoscale and scale-to-zero face the cold start problem, resulting in some initial latency before resolving bursts of unpredicted events. In this context, high-priority events that require to be processed within an established timeframe should not consider scale-to-zero as an option. {\color{black} Since the experiments compared Serverless scenarios with under and over-provisioned configurations, the findings are not limited to the choice of Knative as the Serverless platform. Similar results are expected regardless of the selected platform. }

%% file: 6-conclusion.tex
\section{Conclusion}
\label{section:conclusion}

In this research article, we propose \textit{KTWIN}, a novel Serverless Digital Twin Platform based on Kubernetes. \textit{KTWIN} provides higher-level data model abstractions through open standards and allows \textit{DT} owners and developers to configure details of the underlying services and infrastructure, such as resource allocation, container-based service deployments and auto-scaling policies, without requiring deep technical knowledge. In this context, they can focus their time on defining the system ontologies and writing the application container logic, while the system takes care of deploying and scaling services according to the demand. The \textit{KTWIN} benefits include reduction of operational costs by automating manual operations, more control and flexibility over the underlying container-based services that comprise the Digital Twin landscape, and deployment flexibility to any Kubernetes cluster (on-premise, in the edge, in the cloud) reducing the chance of vendor lock-in.

\textit{KTWIN} innovates in the Digital Twin research field. To the best of our knowledge, this is the first research to propose the implementation of a Kubernetes-based Digital Twin platform built on top of open standards, such as \textit{DTDL}, and Cloud Native tools. Additionally, \textit{KTWIN} proposal includes the usage of Kubernetes Operators to map domain knowledge relationships, represented in a twin graph, into system requirements that include efficient event routing, deployment of user-defined application functions in container-based images, and event storage. The design concept and the implemented prototype enable the foundation for an open-source Kubernetes-based Serverless Digital Twin platform.

The research experiments were designed using real-world domain data models, previously defined by \cite{dtdl-smart-cities} using \textit{DTDL}, and deployed to a Smart City use case based on New York City public data \cite{new-york-mobile-pole-open-data}. The results show that \textit{KTWIN} provides scalability for the different city sizes of the experiments. Regarding resource utilization efficiency and performance, \textit{KTWIN} achieves resource usage efficiency closer to under-provisioned scenarios with cost savings ranging between 60\% and 80\%, while still providing performance near the over-provisioned scenario. It offers configurable auto-scaling policies on top of \textit{Knative}, including scaling to zero, allowing services to be scaled to a static number of resources or scale them based on the actual demand. Finally, because the designed experiments represent a hypothetical day in a modern city in a simulation period of 24 minutes, a 60 times shorter period compared to a full 24-hour day, we can conclude that the implemented prototype should present similar results for cities 60 times larger than our largest simulation scenario. The experiment results and the prototype implemented are publicly available on GitHub \footnote{\url{https://github.com/Open-Digital-Twin/ktwin-article}}.

While the proposed solution offers a robust and flexible tool for enabling Digital Twin use cases, it also presents some limitations. The experiments show that the \textit{KTWIN} \textit{Event Broker} processed up to 1800 requests per second without any impact on message delivery and resource issues, while Knative dispatchers represent a bottleneck of 350 requests per second due to their lack of support for auto-scaling and a fixed amount of allocated CPU and memory. Additionally, in scenarios where scale to zero was configured, services experienced up to 1s-2s to start up, representing some risks for time-sensitive use cases. Time-sensitive workloads can still run on \textit{KTWIN}, but they should not consider auto-scaling to zero as an option if the cold start time impact will not fit into the use case time requirements.



{\color{black}

To further enhance the capabilities of our platform, future research and development efforts will address some of the limitations identified and new requirements that were not in the initial proposed design. The lack of scalability in dispatcher components, representing a bottleneck for the system, will be addressed by implementing multiple consumers for the same event type published to the \textit{Event Broker}. In addition, future research includes the evaluation of \textit{KTWIN} in Edge deployment scenarios and how to distribute the control and application plane components in case the solution is deployed to Cloud and Edge locations. Future works can also evaluate the modeling and deployment of different Digital Twin use cases scenarios such as Manufactoring, Energy, Healthcare or other Smart Cities scenarios. Finally, future works can include Digital Twin Capabilities Table (\textit{CPT}) requirements not tackled in this study such as provide advanced 3D visualization of domain-related data in charts, graphs or dashboards, enable Data Analysis and Analytics using collected event data, implement use cases levering Machine Learning and Edge AI foundation requirements within the platform, and research and validate requirements related to data governance and data privacy. The implemented prototype and deployment setup are available at \url{https://github.com/Open-Digital-Twin}.

}


By providing a flexible and open-source environment for Digital Twin's model and service definitions, we have laid a robust foundation for future innovations. \textit{KTWIN} leverages the rich environment of Kubernetes and we expect that it makes it possible for further enhancements and open collaboration within the \textit{DT} industry and academy, contributing to the advancement of DT technology, driving more intelligent and efficient systems across various domains.

\section*{Acknowledgments}

This research is part of the \textit{INCT of Intelligent Communications Networks and the Internet of Things (ICoNIoT)} funded by CNPq Process 405940/2022-0 and CAPES Finance Code 88887.954253/2024-00. This study was also partially funded by CAPES Finance Code 001.